\documentstyle[aps,prb,twocolumn,overcite,psfig]{revtex}
\begin{document}
\draft

\title{Microscopic Theory of Protein Folding Rates.II: Local Reaction 
Coordinates and Chain Dynamics}

\wideabs{
\author{ John J. Portman \footnotemark[1],
Shoji Takada \footnotemark[2], and Peter G. Wolynes \footnotemark[1]\footnotemark[3]}
\address{\footnotemark[1] Departments of Physics and Chemistry, 
	University of Illinois, Urbana, Illinois, 61801\\  }
\address{\footnotemark[2] Department of Chemistry, 
Kobe University, Rokkodai, Kobe, 657 Japan\\}
\address{\footnotemark[3] Current address: Department of Chemistry and Biochemistry, 
University of California at San Diego, La Jolla, California 92093}

\date{August 23, 2000}
\maketitle
 
\begin{abstract}
{The motion involved in barrier crossing for protein folding are 
investigated in terms of the
chain dynamics of the polymer backbone, completing the microscopic description of protein
folding presented in the previous paper. 
Local reaction coordinates are identified as collective
growth modes of the unstable fluctuations about the saddle-points in the free energy
surface.  The description of the chain dynamics incorporates internal 
friction (independent of the solvent viscosity) 
arising from the elementary isomerizations of the backbone dihedral angles.
We find that the folding rate depends linearly on the solvent friction
for high viscosity, but saturates at low viscosity because of internal friction.
For $\lambda$-repressor,
the calculated folding rate prefactor, along with the free energy barrier from
the variational theory,
gives a folding
rate that agrees well with the experimentally determined rate under highly stabilizing 
conditions, but the theory predicts too large a folding rate at the transition midpoint.
This discrepancy obtained using a fairly complete quantitative theory inspires a 
new set of questions about chain dynamics, 
specifically detailed motions in individual contact formation.
}
\end{abstract}
}
\section{Introduction}
A complete description of protein folding kinetics must
account for the microscopic dynamics of the polypeptide chain.
The relaxation of a free polymer is well described by
Langevin dynamics with an harmonic potential between
adjacent monomers and Markovian random forces \cite{doi_edwards:86}.  
This Rouse-Zimm formalism can be extended through frictional forces
with memory to effectively describe chain motions that include 
many other strictly microscopic complexities such as chain 
stiffness \cite{zwanzig:74,mb:rz:78}.
To describe the chain motions involved in protein folding, one must also
attend to the residue specific interactions or free energy landscape
which guides the protein to its native state and
governs the probability of different ensembles of structures 
described in terms of order parameters.

For many proteins,
the only thermodynamically distinct states with significant 
detectable populations are the unfolded ensemble and the native folded
conformations \cite{jackson:98}.  
This observation suggests the existence of a
free energy barrier  separating
the stable and metastable phases. 
Like nucleation in phase transitions, 
the folding kinetics for these proteins is 
largely governed by the probability of rare configurations 
corresponding to the top of the 
barrier, the so-called transition state ensemble. 
In theoretical studies, the transition state ensembles correspond
with saddle-points in a multi-dimensional free energy 
surface as a function of local order parameters.  
Near the transition state, some order parameter fluctuations
grow to become the folded phase.  
This is an unstable mode of collective motion.  Motion in 
other directions tend to stabilize the transition state.  
The structure of the unstable growth mode 
can be thought of as a locally defined reaction coordinate.
If there are multiple saddle points which are connected,
the local reaction coordinate for each of these will usually differ.
In such cases, especially, the local reaction coordinate may not
be the best way of describing the global transformation of the
ensemble involved in folding.

Lately, there has been a great effort to elucidate the
nature of the transition state ensemble and ideal reaction coordinate
for specific proteins through measurement 
\cite{lsi:deo:arf:95,elh:ls:96,bhdco:97a,ngnssf:97,%
grsd:98,jcm:mtp:ls:98}, 
theoretical calculations 
based on physically intuitive choices
\cite{bas:jw:pgw:99,jjp:st:pgw:98,ovg:avf:99,ea:db:99,vm:wae:99},
and simulations \cite{via:amg:eis:94,oslw:96,%
shakhnovich:97,ajl:vd:96,%
dkk:dt:98,js:jno:clb:99,vsp:dsr:99}. 
While many aspects of the folding kinetics seem to be understood,
still some basic issues remain controversial.
For example, while the structural distribution of the transition state 
ensemble is well predicted 
\cite{bas:jw:pgw:99,jjp:st:pgw:98,ovg:avf:99,ea:db:99,vm:wae:99}, 
the absolute magnitude of the barrier height is not well determined.
Comparison with experiment is clouded by the uncertainty 
of the prefactor of the folding rate. 

For simple thermally activated transitions, one expects
the rate to follow the activated law
\begin{equation}\label{eq:rate}
k = k_0 e^{- \beta \Delta F^{\dagger}},
\end{equation}
where $\beta = 1/k_B T$ is the inverse temperature 
and $\Delta F^{\dagger}$ is the barrier height which contains both
entropy and energy.
For folding, $\Delta F^{\dagger}$ is expected to be strongly 
temperature dependent.
The exponential factor is the relative probability of finding the system 
at the transition state compared to the metastable minima.  
The prefactor, $k_0$, is determined by the microscopic dynamics 
involved in crossing
the barrier.  For proteins with very rugged landscapes,
the dynamics can be described as hopping between trapped states 
\cite{jdb:pgw:89}.
In this case the barriers to escape traps enter strongly.
For proteins with very little frustration that have 
smoother landscapes, there will be little trapping, but
the chain motion will be diffusive and depend on chain connectivity.
$k_0$ depends on the free energy surface and the frictional forces
such as viscosity that determine the stochastic motion of the protein
as it folds. 
One estimate proposed for the prefactor relating $k_0$
to the inverse time it takes to form a typical loop given by simple 
polymer theory, 
$k_0 \approx 1\mu \mathrm{s}^{-1}$ \cite{zg:dt:95,hhse:96}.
While quite reasonable,
this estimate does not involve the shape of the
free energy surface near the transition state.
Thus, for example, it is not completely clear how big a loop should be 
considered.

In Ref. \citen{jjp:st:pgw:98} and the preceding paper
(henceforth referred as [I]), we presented a variational theory for
the protein folding free energy surface of a specific protein
to a known structure.  In this approach, the protein is modeled as
a collapsed, stiff chain with interactions between residues in 
contact in the native
structure.  The reference Hamiltonian is that of a polymer chain
inhomogeneously constrained to the native structure.
The free energy profile for folding obtained from  model can
be described as an activation barrier between the 
unfolded and folded ensembles
with fine structure from local minima and saddle-points in the free 
energy surface giving rise to longitudinal ruggedness in the free 
energy profile for a perfectly funneled landscape.
In this approach, conformational ensembles 
along the folding route are characterized by structural 
order parameters evaluated
as averages over the distribution from the reference Hamiltonian.
The focus of the present paper is how the microscopic 
chain dynamics are involved in barrier crossing within this model.

Understanding barrier crossing dynamics in protein folding requires not only
a good characterization of the free energy surface and structural ensembles,
but also a good description of the microscopic motions of the polymer backbone.
In the Rouse-Zimm model of polymer dynamics, the 
motion of the polymer chain is governed by the chain connectivity, represented
as harmonic entropic springs connecting adjacent monomers, and is damped by the 
interactions of the monomers with the solvent \cite{doi_edwards:86}.
A more detailed description of polypeptide dynamics acknowledges there 
are activated transitions between 
{\em gauche} and {\em trans} bond angles determined by the
dihedral potential 
\cite{helfand:71,js:eh:80,eh:js:82}.
One way to incorporate the short range structural
information arising from these barriers to internal rotation
is through a static chain stiffness that defines renormalized
harmonic correlations 
\cite{mb:rz:78,ga:fg:81,ap:sb:cc:84,ap:mg:86,ap:fg:ga:87,perico:89,%
perico:89b,hffp:91,ap:nem:mde:98}.  
However, a complete description must also incorporate the 
activated nature of these rotational isomerization rates in the 
dynamical correlations.

Quite early in the study of polymer dynamics, anomalies found
in intrinsic viscosity measurements indicated that there was a
source of friction independent of the viscosity of the solvent 
\cite{kuhn2:46,cerf:57}.  
Kuhn and Kuhn first suggested this ``internal viscosity'' was a 
consequence of microscopic barrier hopping in the local chain 
segments \cite{kuhn2:46}. 
This phenomenological friction found for high polymers may 
have a variety of microscopic origins in addition to internal 
friction due to barrier hopping of many elementary 
chain links.
A helpful (though perhaps dated) evaluation of different theories can be found
found in Ref. \citen{cwm:mcw:85}.  
Fixman  has shown that simulations
for polymers with rigidly fixed bond angles and bond lengths 
can be mimicked by internal viscosity \cite{fixman:87,fixman:89}.
Still other possible sources of the measured internal 
viscosity include the effect of interactions with distant monomers in the 
chain \cite{deGennes:79}, or other inherent non-linearities neglected in the 
Gaussian model assumption \cite{saa:kff:77}. To our knowledge, the microscopic 
interpretation of internal viscosity for artificial high polymers
has not been definitively settled. 
It is clear, however, that polypeptides will at least exhibit internal
viscosity from the difficulty of making backbone angle transitions.  
These transitions are roughly 100 times slower than if no barriers 
were present in the Ramachandran potential.

In this paper, we complete the logical development of folding rate
theory for minimally frustrated proteins by 
analyzing the microscopic dynamics of chain motions involved in
barrier crossing of the protein folding model presented 
in Ref. \citen{jjp:st:pgw:98} and [I].
We use multi-dimensional, non-Markovian Langevin dynamics to describe 
the barrier crossing, and identify the unstable modes as the reaction 
coordinates. For the folding rate calculation, we use a 
generalization of Kramer's theory \cite{kramers:40} that treats
non-Markovian dynamics in many dimensions \cite{amb:ep:vyz:92}.  
At the same time, the interpretation of a local 
reaction coordinate follows
Langer's theory of the nucleation rates \cite{langer:69}.
The details of chain dynamics enter the formalism through a generalized friction 
matrix with memory.  The friction that damps the motion of the chain
includes the solvent viscosity and an internal friction 
that depends on the length scale (in sequence)
of the motion but which we take as is independent 
of the solvent viscosity. 
The internal friction accounts for the activated nature of dihedral angle
flips in polypeptides.
We restrict our attention to over-damped 
dynamics, though it is easy to generalize the calculations to include 
inertial terms as well.  

\section{Reaction Coordinates and Folding Rates}
To calculate the barrier crossing rate, one must first identify the
order parameters that describe the phases of the system, and then
study their motions. In density functional theory, one
often considers a density field that is slowly varying in space, and 
expands the density functional about a uniform density \cite{oxtoby:91}.
This expansion yields a free energy consisting of a bulk free energy
and can be expanded in gradient terms that ultimately represent
the surface tension between two phases.
The density then evolves either hydrodynamically or diffusively under
the influence of the free energy driving force.
The quasi-equilibrium part of this approach has been applied to 
nucleation of the folded state in proteins
starting with a mean field free energy functional 
and defining a global native density field
as a measure of similarity to the native state \cite{st:pgw:97a}
\begin{equation}\label{eq:natdens1}
\rho^{{\rm N}}({\bf r},\{{\bf r}_i\}) 
\sim \sum_i 
\delta( {\bf r} - {\bf r}_i) \delta ({\bf r_i} - {\bf r}_i^{\rm N}),
\end{equation}
where $\{{\bf r}_i\}$ specify the configuration of the protein chain, 
and $\{{\bf r}_i^{\rm N}\}$ are the corresponding coordinates 
of the native structure.
In the present work, we wish to have a description
of the folding that is local in sequence
and hence consider each term in the sum separately.
This local native density defined as
\begin{equation}\label{eq:loc_ndens2}
\rho^{{\rm N}}({\bf r},{\bf r}_i) = \delta({\bf r} - {\bf r}_i) 
\exp \left[
-\frac{3}{2a^2} \alpha^{{\rm N}}({\bf r}_i - {\bf r}_i^{{\rm N}})^2
\right],
\end{equation}
where we have relaxed the measure of similarity of the native state 
by replacing
$\delta ({\bf r_i} - {\bf r}^{\rm N})$ 
by a Gaussian measure.

To connect this density to our folding route calculations, 
it is more convenient to consider scalar order parameters than to 
expand the free energy in terms of field variables. 
Therefore, we integrate $\rho({\bf r},{\bf r}_i)$ over ${\bf r}$,
to give the scalar measure native similarity
\begin{equation}\label{eq:loc_ndens}
\rho({\bf r}_i) = 
\exp \left[
-\frac{3}{2a^2} \alpha^{{\rm N}}({\bf r}_i - {\bf r}_i^{{\rm N}})^2]
\right].
\end{equation}
[To simplify notation, we have suppressed the superscript $N$ 
since the native density can still be distinguished by the 1-body monomer 
density, $\rho^{1}({\bf r})$, given in 
Eq.(I-25)]. 
Following the interpretation of the order parameters given in [I], 
we calculate the average native density of site $i$,
$\rho_i = \langle \rho({\bf r}_i) \rangle_0$,
for a given set of variational constraints 
$\{C_i\}$
\begin{eqnarray}\label{eq:rhoave}
\lefteqn{\rho_i[\{C\}]  =  \int \! \! d{\bf r} \: \rho_i^{1}({\bf r}) \rho({\bf r})} \hspace{.5cm}\\
& &
= (1 + \alpha^{{\rm N}} G_{ii})^{-3/2} 
\exp 
\left[ - \frac{3}{2a^2} 
\frac{\alpha^{{\rm N}}({\bf s}_i - {\bf r}_i^{{\rm N}})^2 }
{1 + \alpha^{{\rm N}} G_{ii}} 
\right] ;
\end{eqnarray}
that is, $\rho_i$ is a function of $\{C_i\}$ 
through the correlations, $G_{ij}$, and average position, ${\bf s}_i$,
given by 
Eq.(I-21) 
and 
Eq.(I-22), 
respectively.
The free energy can then be parameterized by $\{\rho_i\}$ with
${\bar F}[\{\rho\}] = F[\{C\}]$ where
$\{\rho\}$ is evaluated at $\{C\}$.  
In particular, we denote a saddle-point of  $F[\{C\}]$ by $\{C^{\star}\}$
and the corresponding native density by 
$\rho_i^{\star} = \rho_i[\{C^{\star}\}]$.

We assume Langevin dynamics with memory for the native density,
$\{\rho_i(t)\}$,
\begin{equation}\label{eq:rho_gle1}
\partial_t \rho_i(t) = 
- \sum _j \int_0^t \! dt' \mu_{ij}(t-t') 
\frac{\partial \beta {\bar F}[\{ \rho(t') \}]}{\partial \rho_j}
+ \xi_i(t),
\end{equation}
where $\beta = 1/k_B T$ is the inverse temperature, and 
$\mu(t)$ is a generalized mobility matrix related to the random noise
$\xi_i(t)$ through the correlations
$\langle \xi_i(t) \xi_j(t') \rangle = \mu_{ij}(t-t')$.
The form of $\mu(t)$ is discussed
in the following section, here we assume that $\mu(t)$ is a known
function of time.  

To study the dynamics near the saddle-point, we 
expand $\bar{F}[\{\rho\}]$ about 
$\rho_i^{\star}$ to second order
\begin{equation}\label{eq:fexpansion}
\beta \bar{F}[\{\rho\}] \approx
\mathrm{const} + \frac{1}{2} \sum_{ij} {\bar \Gamma}_{ij}^{\star} 
\delta \rho_i \delta \rho_j,
\end{equation}
where 
$\delta \rho_i = \rho_i - \rho_i^{\star}$,
and the Hessian matrix
\begin{equation}\label{eq:Gammabar}
{\bar \Gamma}_{ij}^{\star}
= 
\frac{\partial^2 \beta {\bar F}[\{\rho^{\star}\}]}
{\partial \rho_i \partial \rho_j}
\end{equation}
has one negative eigenvalue since it is evaluated at a saddle-point.
In terms of derivatives with respect to the variational parameters,  
${\bar \Gamma^{\star}}$ is the solution of the matrix equation
\begin{equation}\label{eq:chainrule}
\sum_{kl}
\bar \Gamma_{kl}^{\star}
\frac{\partial \rho_k^{\star}}{\partial C_i}
\frac{\partial \rho_l^{\star}}{\partial C_j} = 
\frac{\partial^2 \beta F[\{C^{\star}\}]}{\partial C_i \partial C_j}.
\end{equation}
where the derivatives with respect to $C_i$ can calculated as 
described in [I].

With this approximation, Eq.(\ref{eq:rho_gle1}) can be written as
\begin{equation}\label{eq:rho_gle}
\partial_t (\delta \rho_i(t)) = 
- \sum_j \int_0^t \! dt' \left[ \mu(t-t'){\bar \Gamma}^{\star} \right]_{ij}  
\delta \rho_j(t') + \xi_i(t).
\end{equation}
Accordingly, the native density correlation functions 
${\cal C}(t)$ with components
\begin{equation}\label{eq:crho_gle}
{\cal C}_{ij}(t) = \langle \delta \rho_i(t) \delta \rho_j(0) \rangle_0,
\end{equation}
satisfy 
\begin{equation}
\partial_t {\cal C}(t) = 
- \int_0^t \! dt' \mu(t-t'){\bar \Gamma}^{\star} {\cal C}(t').
\end{equation}
In Eq.(\ref{eq:rho_gle}), we have subtracted the stationary value
$ \rho_i(t \rightarrow \infty) = \rho_i^{\star}$, so that 
$\delta \rho_i(t \rightarrow \infty) = 0$ and 
${\cal C}(t \rightarrow \infty) = 0$.
This long time solution is unstable, however, because it is a 
saddle-point of the free energy. 

The unstable mode can be determined by the average equation of 
motion \cite{langer:69}.
Denoting the Laplace transform of an arbitrary function of time
by $\hat{g}(\omega) = \int_0^\infty \! dt e^{-\omega t} g(t)$,
the solution of the Eq.(\ref{eq:rho_gle}) averaged over the random
noise is
\begin{equation}\label{eq:lt_averho}
\overline{\delta \hat{\rho}_i(\omega)} = 
\left[ \omega {\bf 1} 
+ \hat{\mu}(\omega) \bar{\Gamma}^{\star} \right]^{-1}_{ij} 
\delta \rho_j(0).
\end{equation}
By inverting the Laplace transform, $\overline{\rho_i(t)}$ can be 
expressed as a sum of exponentials with time constants determined 
by the poles of
$\left[ \omega {\bf 1} 
+ \hat{\mu}(\omega) \bar{\Gamma}^{\star} \right]^{-1}$: 
the eigenfunctions
$\hat{\mu}(-\kappa) \bar{\Gamma}^{\star} \cdot {\bf u} 
= \kappa {\bf u}$
have time dependence ${\bf u}(t) \sim e^{-\kappa t}$. Assuming that
$\hat{\mu}(\omega)$ is positive definite, there is one mode for which
$\kappa$ is negative, because $\bar{\Gamma}$ is the 
saddle point Hessian of $F[{\rho}]$:
\begin{equation}\label{eq:usmode_t}
{\bf u}^{\star}(t) = {\bf u}^{\star} e^{\left| \kappa^{\star} \right| t},
\end{equation}
where 
\begin{equation}\label{eq:usmode}
\hat{\mu}(-\kappa^{\star}) \bar{\Gamma}^{\star} \cdot {\bf u}^{\star}_{j} = 
\kappa^{\star} {\bf u}^{\star}_{j} 
\; \; \; \; \; \; \;
(\kappa^{\star} < 0).
\end{equation}
Since motion along the unstable mode ${\bf u}^{\star}$ grows 
exponentially away from the saddle point, 
we identify the  components
of  ${\bf u}^{\star}$ as the local reaction coordinate to
surmount the barrier.  Both ${\bf u}^{\star}$ and 
$\kappa^{\star}$ are essentially the unstable mode and curvature of 
$\bar{F}[{\rho}]$, but renormalized by the dynamics of the barrier crossing 
incorporated in the mobility matrix. In Eq.(\ref{eq:usmode}), 
$\kappa^{\star}$ is the multi-dimensional analogue of the Grote-Hynes 
frequency (in the over-damped limit) \cite{rfg:jth:80}.

The rate for barrier crossing corresponding to Eq.(\ref{eq:rho_gle})
is \cite{amb:ep:vyz:92}
\begin{equation}\label{eq:rate1}
k = \frac{|\kappa^{\star}|}{2 \pi} 
\left| \frac{\det \bar{\Gamma}_{MS}}{\det \bar{\Gamma}^{\star}} \right|^{1/2}
e^{-\beta \Delta F^{\dagger}},
\end{equation}
where $\Delta F^{\dagger}$ is the barrier height and
$\bar{\Gamma}_{MS}$ and $\bar{\Gamma}^{\star}$ are the curvature 
matrices of $\bar{F[{\rho}]}$ evaluated at the metastable minimum and 
the saddle-point, respectively.  Eq.(\ref{eq:rate1}) a generalizes both 
the rate calculations presented by Langer \cite{langer:69} and 
Grote-Hynes \cite{rfg:jth:80}; it simultaneously accommodates both 
multi-dimensional diffusion (as in the Langer formula) as well as time 
dependent friction (as in the Grote-Hynes formula). 

The ratio of determinants accounts for the entropic differences 
between the transition state and the metastable phase
due to fluctuations of the order parameter.  
These fluctuations are like the capillary
waves which renormalize the surface tension in ordinary nucleation.
These contributions would already be included in an exact free energy 
so that the determinants should be absorbed into the exponential 
factor to most simply keep a consistent level of
thermodynamic theory \cite{langer:73}.
Consequently, the expression for the rate becomes
\begin{equation}\label{rate}
k = \frac{|\kappa^{\star}|}{2 \pi} e^{-\beta \Delta F^{\dagger}},
\end{equation}
where $\kappa^{\star}$ is given by Eq.(\ref{eq:usmode}).
As pointed out in Ref. \citen{dfc:pgw:83}, if the over-counting of 
entropy were not corrected, the ratio of the forward and backwards 
rate would not equal the equilibrium constant
as determined by the starting free energy functional, 
$k_{12}/k_{21} \ne e^{-\beta(F_2 - F_1)}$.

\section{Linear Response Approximation of the Generalized Mobility}

The time evolution of the 
native densities $\{\delta \rho_i(t)\}$ depends on 
microscopic dynamics of the monomer positions $\{{\bf r}_i(t)\}$.  
The chain dynamics, in turn, are incorporated into the effective mobility 
matrix $\mu(t)$ in the relaxation equation for $\delta \rho(t)$ 
(Eq.(\ref{eq:rho_gle})).  
To determine $\mu(t)$
consistent with the definition of the native densities this theory, 
we consider the chain dynamics near the saddle point to 
be governed by the $H_0$ with constraints $\{C^{\star}\}$.
The resulting mobility matrix for native densities in Eq.(\ref{eq:rho_gle}) 
is assumed to be same as that which arises from the Mori-Zwanzig 
projected dynamics \cite{zwanzig:61,mori:65}.
We begin  by outlining this formalism as it was 
applied in Zwanzig's classic paper on generalized Rouse 
dynamics for a polymer described by an arbitrary 
potential \cite{zwanzig:74}. 
A closed form of the mobility matrix is then obtained by 
approximating the local dynamics with those of the constrained 
polymer Hamiltonian, $H_0$.

Consider a polymer described by the potential $U({\bf R})$
with monomer positions ${\bf R} = ({\bf r}_1, \ldots , {\bf r}_n)$.
The probability density of monomer
positions, $\Psi({\bf R},t)$ is assumed to evolve according to the 
Smoluchowski equation
\begin{equation}\label{eq:smol}
\partial_t \Psi({\bf R},t) = {\cal D} \Psi({\bf R},t)
\end{equation}
\begin{equation}\label{eq:diffop}
{\cal D} = \sum_{ij} 
\nabla_i \cdot D_{ij} e^{-\beta U({\bf R})} \nabla_j e^{\beta U({\bf R})}
\end{equation}
where $\nabla_i \equiv \partial/\partial {\bf r}_i$, and $D$ is the 
diffusion matrix.
In this section, we denote the 
equilibrium averages by 
$\langle \ldots \rangle = \int \! d{\bf R} \: \ldots \Psi_{eq}({\bf R})$
where
\begin{equation}\label{eq:boltzman}
\Psi_{{\rm eq}}({\bf R}) =  e^{-\beta U({\bf R})}
\left[\int \! \! d{\bf R} \: e^{-\beta U({\bf R})}\right]^{-1}.
\end{equation}

Now consider a set of dynamical variables that are 
functions of the monomer positions, $\{A_i({\bf R})\}$.  These variables 
will be chosen to be the order parameters $\{\rho\}$ to 
describe the barrier crossing 
dynamics discussed in the previous section, but for now $\{A\}$ is 
arbitrary.  
The equilibrium time dependent correlation functions of
\begin{equation}\label{eq:deltaA}
\delta A_i = A_i - \langle A_i \rangle
\end{equation}
are denoted by 
\begin{equation}\label{eq:corrAAdef}
{\cal C}^{A}_{ij}(t) \equiv
\langle \delta A_i(t) \delta A_j(0) \rangle.
\end{equation}
The time dependent correlations can be formally expressed as
\begin{eqnarray}\label{eq:corrAA}
{\cal C}^{A}_{ij}(t) & = &
\int \! \! d{\bf R} \delta A_i e^{{\cal D}t} 
(\delta A_j \Psi_{{\rm eq}}) 
\nonumber \\
& = &
\int \! \! d{\bf R} \Psi_{{\rm eq}} \delta A_i e^{{\cal L}t} 
\delta A_j.
\end{eqnarray}
In the second line of Eq(\ref{eq:corrAA}), 
we have introduced the adjoint operator ${\cal L}$ defined by its
action on an arbitrary function $B$
\begin{equation}\label{eq:adjprop}
{\cal D}\Psi_{{\rm eq}}B = \Psi_{{\rm eq}} {\cal L} B,
\end{equation}
or explicitly,
\begin{equation}\label{eq:adjop}
{\cal L} = \sum_{ij}
e^{\beta U({\bf R})} 
\nabla_i D_{ij} e^{-\beta U({\bf R})} \cdot \nabla_j.
\end{equation}

The meaning of the operator notation in Eq.(\ref{eq:corrAA}) is defined
through the propagator (or Green's function) 
$P({\bf X} \: t; {\bf Y})$:
\begin{equation}\label{eq:corrAA_P}
{\cal C}^{A}_{ij}(t) = 
\int \! \! d{\bf X} d{\bf Y} \:
\delta A_i({\bf X}) \delta A_j({\bf Y}) P({\bf X} \: t; {\bf Y}) 
\Psi_{{\rm eq}}({\bf Y}),
\end{equation}
where (formally)
\begin{equation}\label{eq:prop}
P({\bf X} \: t; {\bf Y}) = e^{{\cal D}t}\delta({\bf X}-{\bf Y}),
\end{equation}
i.e.,
$P({\bf X} \: t; {\bf Y})$ satisfies the Smoluchowski equation,
$\partial_t P = {\cal D}P$, 
with initial conditions
$P({\bf X} \: t = 0; {\bf Y}) = \delta( {\bf X} - {\bf Y})$.

An alternative equation can be 
derived for the
correlations through the projected dynamics of Eq.(\ref{eq:smol}).  
Following Ref. \citen{zwanzig:74}, 
we define the projection operator through its action on a arbitrary function
$B$
\begin{equation}\label{eq:proj}
{\cal P}B = \sum_{ij} \delta A_i \Gamma^{{\rm A}}_{ij} 
\langle \delta A_j B \rangle
\end{equation}
where $(\Gamma^{{\rm A}})^{-1}$ is the matrix of static correlations
\begin{equation}\label{eq:gamma_a}
(\Gamma^{{\rm A}})^{-1} = \langle \delta A_i \delta A_j \rangle.
\end{equation}
Using standard projection operator techniques \cite{zwanzig:61,mori:65}, 
it can be shown
that the correlations obey the generalized Langevin equation \cite{zwanzig:74}
\begin{equation}\label{eq:cA_gle}
\partial_t {\cal C}^{\rm A}(t) = - \Omega^{{\rm A}} 
\Gamma^{{\rm A}} {\cal C}^{{\rm A}} + 
\int_0^{t} \! dt' K^{{\rm A}}(t-t') \Gamma^{{\rm A}} {\cal C}^{{\rm A}}(t'),
\end{equation}
where
\begin{equation}\label{eq:omega}
\Omega^{{\rm A}}_{ij} = - \langle A_i {\cal L} A_j \rangle 
= \sum_{kl} \langle \nabla_k A_i \cdot D_{kl} \cdot \nabla_l A_j \rangle
\end{equation}
and $K(t)$ is the memory kernel
\begin{equation}\label{eq:K_memory}
K^{{\rm A}}_{ij}(t) = \langle A_i {\cal L}(1 - {\cal P})  
\exp[ (1 - {\cal P}){\cal L} t](1 - {\cal P}) 
{\cal L} A_j \rangle.
\end{equation}
Eq.(\ref{eq:cA_gle}) can be written in the form of Eq.(\ref{eq:crho_gle})
\begin{equation}\label{eq:cAmu_gle}
\partial_t {\cal C}^{\rm A}(t) = - 
\int_0^{t} \! dt' \mu^{{\rm A}}(t-t') \Gamma^{{\rm A}} {\cal C}^{{\rm A}}(t'),
\end{equation}
where 
\begin{equation}\label{eq:muA}
\mu^{{\rm A}}(t) = 2 \Omega^{{\rm A}} \delta(t) - K^{{\rm A}}(t)
\end{equation}
is the effective time-dependent mobility matrix.  As suggested by this
notation, the static correlations $(\Gamma^{{\rm A}})^{-1}$ can be thought
of as arising from an effective harmonic free energy 
$\beta F = \mathrm{const} + (1/2) \sum_{ij} \delta A_i \Gamma_{ij}^{\mathrm{A}} \delta A_j$ .

Eq.(\ref{eq:cAmu_gle}) is an exact result, 
equivalent to Eq.(\ref{eq:corrAA_P}).
The linear response formulation  avoids 
explicit use of the Green's function which is generally unknown
for an arbitrary potential. 
On the other hand, it is very difficult to calculate the 
memory kernel explicitly from Eq.(\ref{eq:K_memory}) since the 
projection operators are very awkward to manipulate.

If we choose $\{A\}$ to be the monomer positions, then
Eq.(\ref{eq:cA_gle}) becomes an equation for the monomer correlation.
If one neglects the memory term, the monomer
correlations are said to obey
optimized Rouse dynamics \cite{mb:rz:78,zwanzig:74}.  
This is equivalent to having
an effective pre-averaged diffusion matrix, $\Omega = \langle D \rangle$,
and a Gaussian approximation to the chain through the static correlations.
[Bixon and Zwanzig first presented the stiff chain connectivity 
matrix 
[Eq.(I-4)] 
in a paper that employs the optimized Rouse dynamics formalism].
One usually must resort to approximate methods to include 
the memory kernel in the analysis.
A clear presentation of these methods applied to one-dimensional problems 
is given in Ref. \citen{ppfps:93}.

One way to study the memory function for monomer correlations is through
a truncated Mori continued fraction \cite{grigolini:85}. This becomes 
rapidly more cumbersome as the order of the expansion is increased.
A more convenient expansion of the memory kernel is to first expand 
the Green's function in Eq.(\ref{eq:corrAA_P}) in the eigenfunctions 
of ${\cal L}$, and then expand these eigenfunction in a finite basis 
(similar to what is done in quantum chemistry calculations)
\cite{hmcffp:90,xyc:kff:93,wht:xyc:kff:95,ksk:kff:97,%
ap:rp:97,flcp:99,glp:rp:ap:99}.  
The basis functions must be chosen with some care for accurate results.
Furthermore, in both the continued fraction and eigenfunction expansion 
methods require equilibrium averages that are difficult to compute for an 
arbitrary potential, requiring simulations in general.  Nevertheless, 
this is a useful method since the simulation time required for the equilibrium 
averages can be much shorter than would be required to simulate the 
time-dependent correlation functions themselves \cite{xyc:kff:93}.

To connect to the barrier crossing calculation for protein folding,
we need to study the dynamics of the native density, $\{A\} = \{\rho\}$.
For a general potential, calculating the correlation functions 
$\langle \delta \rho_i(t) \delta \rho_j(0) \rangle$ involves
the same technical difficulties that arises for the calculation of the
monomer correlations.
We avoid these complications by approximating the potential by 
the constrained polymer Hamiltonian, $H_0$. 
Not only is this is a reasonable choice given the 
interpretation of the order parameters as averages over $H_0$, it
also allows us to calculate the correlation from Eq.(\ref{eq:corrAA_P})
directly.

For an harmonic potential, the propagator in Eq.(\ref{eq:corrAA_P}) can
be solved analytically \cite{risken:89}.
The Gaussian form of $\rho({\bf r}_i)$
allows the integrals in Eq.(\ref{eq:corrAA_P}) to be performed to give
the native density correlations (see Appendix for details)
\begin{eqnarray}\label{eq:rhocorr}
\langle \rho_i(t) \rho_j(0) \rangle_0 &= &
(\det M_{ij}(t))^{-3/2} \nonumber \\
&\times&
\exp \left[\frac{- 3 \alpha^N}{2 a^2}
{\bf J}_{ij}^T \cdot M_{ij}(t)^{-1} 
{\bf J}_{ij}
\right]
\end{eqnarray}
where 
\begin{equation}\label{eq:def_SM}
M_{ij}(t) = 
\left[ \begin{array}{cc}
(1 + \alpha^{{\rm N}} G_{ii}) & \alpha^{\rm N}G_{ij}(t) \\
\alpha^{\rm N} G_{ij}(t) & (1 + \alpha^{{\rm N}} G_{jj})
\end{array} \right],
\end{equation}
and, ${\bf J}_{ij}$ is the 2-component difference vector
\begin{equation}\label{eq:Sij}
{\bf J}_{ij} = 
\left[
\begin{array}{c}
{\bf r}^N_i - {\bf s}_i \\
{\bf r}^N_j - {\bf s}_j 
\end{array}
\right].
\end{equation}
In this expression, we have introduced the time 
correlation function of monomer positions
\begin{equation}\label{eq:gt_def}
G_{ij}(t) = 
\langle \delta {\bf r}_i(t) \cdot \delta {\bf r}_j(0) \rangle_0/a^2.
\end{equation}
The chain dynamics are contained in $G_{ij}(t)$, and the precise form
depends on the diffusion matrix $D_{ij}$ in Eq.(\ref{eq:diffop}).  
We postpone further discussion of $G_{ij}(t)$ until the next section.

The correlations in Eq.(\ref{eq:corrAA_P}) for $\{A\} = \{\rho\}$ are written
as 
\begin{equation}\label{eq:drhocorr}
{\cal C}_{ij}(t) = \langle \delta \rho_i(t) \delta \rho(0) \rangle_0 
= \langle \rho_i(t) \rho_j(0) \rangle_0 - \rho_i \rho_j
\end{equation}
where the first term is given by Eq.(\ref{eq:rhocorr}-\ref{eq:Sij}) and
the second term is found from Eq.(\ref{eq:rhoave}).

Since approximating the chain potential with $H_0$ permits an exact 
solution for the correlation functions, it may seem as if the discussion 
of the projected dynamics was an unnecessary detour. However, what we 
are really trying to calculate is the effective mobility matrix needed 
to evaluate the folding rate prefactor in Eq.(\ref{eq:usmode}).  

From Eq.(\ref{eq:cAmu_gle}), the native density correlations obey
the projected relaxation equation
\begin{equation}\label{eq:rhocorr_glep}
\partial_t {\cal C}(t) = - 
\int_0^{t} \! dt' \mu^{\rho}(t-t') \Gamma^{\rho} {\cal C}(t'),
\end{equation}
where $\Gamma^{\rho}$ are the inverse of the static correlations
\begin{equation}
[\Gamma^{\rho}]^{-1}_{ij} = \langle \delta \rho_i \delta \rho_j \rangle_0
= {\cal C}_{ij}(0).
\end{equation}
Eq.(\ref{eq:muA}) with  $\{A\} = \{\rho\}$ gives $\mu^{\rho}(t)$, in terms
of projection operators, but we can also deduce an alternative 
expression for the mobility since ${\cal C}(t)$ is known.
Solving Eq.(\ref{eq:rhocorr_glep}) for $\mu^{\rho}(t)$ by Laplace transforms gives
\begin{equation}\label{eq:mu_rho}
\hat{\mu}^{\rho}(\omega) = {\cal C}(0) \hat{{\cal C}}(\omega)^{-1} {\cal C}(0)
- \omega {\cal C}(0)
\end{equation}
To close this expression for the mobility, we only need the Laplace transform 
of ${\cal C}(t)$ which can be calculated numerically.

As discussed previously, Eq.(\ref{eq:rhocorr_glep}) is consistent
with a Gaussian approximation to the free energy 
\begin{equation}\label{eq:frho_p}
\beta F^{\rho} = {\rm const} + \frac{1}{2}
\sum_{ij} \Gamma^{\rho}_{ij} \delta \rho_i \delta \rho_j
\end{equation}
analogous to the saddle-point expansion of $\bar{F}[\{\rho^{\star}\}]$ 
given in Eq.(\ref{eq:fexpansion}).  However, even though $\Gamma^{\rho}$ 
can be evaluated at the saddle-point constraints in $\{C^{\star}\}$ 
in $H_0$, $F^{\rho}$ is stable against all fluctuations away 
from $\delta \rho_i = 0$.  This is in contrast to the curvature 
matrix $\bar{\Gamma}^{\star}$ that has one unstable direction.

Nevertheless, by approximating the mobility in Eq.(\ref{eq:crho_gle})
by $\mu(t) \approx \mu^{\rho}(t)$ we
can solve the eigenvalue equation Eq.(\ref{eq:usmode}) to identify
the unstable mode ${\bf u}^{\star}$ and calculate the prefactor 
$\kappa^{\star}$.  The last remaining unspecified quantity for the 
theory is the time dependent monomer correlations $G(t)$ which is 
the subject of the next section.

%
\section{Chain Dynamics}

In this section we specify our model for the chain dynamics, and 
calculate the monomer pair correlation, $G(t)$, with the approximation
to the chain potential $U({\bf R}) = H_0$. This Gaussian approximation 
simplifies the true microscopic potential that governs 
conformational changes of the protein chain, and by itself ignores
explicit interactions.  We attempt to capture some effects of the 
microscopic dynamics by including frictional forces that are independent 
of the solvent viscosity, i.e., internal viscosity.

\subsection{Monomer Pair Correlations}
Starting from the definition of pair correlations for the Smoluchowski equation
[Eq.(\ref{eq:corrAA_P})], it is easy to derive an equation for the 
relaxation of the monomer correlations 
\begin{equation}\label{eq:gt1}
\partial_t \langle \delta {\bf r}_i(t) \cdot \delta {\bf r}_j(0) \rangle
=  \sum_k \gamma_{ik}^{-1} 
\langle \nabla_k U(\{{\bf r}(t)\}) \cdot \delta {\bf r}_j(0) \rangle,
\end{equation}
where the friction coefficient is related to the diffusion coefficient by
\begin{equation}\label{eq:gamma_D}
\beta D = \gamma^{-1}.
\end{equation}
In Eq.(\ref{eq:gt1}), we have assumed that $\gamma$ is independent of 
$\{{\bf r}\}$, but may still  depend on the sequence index.  
The model for the friction matrix is 
specified shortly, for now we define a dimensionless friction matrix, 
$\tilde{\gamma}$:
\begin{equation}\label{eq:gamma_0}
\gamma = \gamma_0 \tilde{\gamma}
\hspace{1cm}
\gamma_0 = 6 \pi a_{{\rm eff}} \eta,
\end{equation}
where we have used Stokes law to define $\gamma_0$ with 
the solvent viscosity $\eta$ and typical monomer van der Waals radius
$a_{{\rm eff}}$.

For a given point along the folding route (e.g., a transition state),
we assume that the chain dynamics can be described by the reference 
Hamiltonian.  
[Eq.(I-13)]. 
Physically, $H_0$ corresponds to a polymer confined to the
native structure by harmonic constraints with spring constants 
$\{3/2a^2 C_i\}$.  These constraints act as a non-uniform external 
field, controlling the fluctuations about the native structure.
Taking the gradient of $H_0$, and using the definition 
[Eq.(I-15)] 
of the average monomer position, ${\bf s}_i$,  gives
\begin{eqnarray}\label{eq:gradH0}
\frac{\partial H_0}{\partial {\bf r}_i} & = & 
\frac{3 k_B T}{a^2}
\left[ \sum_j \Gamma_{ij}^{{\rm (ch)}}{\bf r}_j 
+ C_i ({\bf r}_i - {\bf r}_i^{{\rm N}}) \right]
\nonumber \\
& =  & \frac{3 k_B T}{a^2} \sum_j \Gamma_{ij}^{(0)} \delta {\bf r}_j,
\end{eqnarray}
where $\Gamma_{ij}^{(0)}$ is given in 
Eq.(I-17) 
and $\delta {\bf r}_i = {\bf r}_i - {\bf s}_i$ as usual.
Substituting this into Eq.(\ref{eq:gt1}) gives the matrix equation
for the correlations
$G_{ij}(t) = \langle \delta {\bf r}_i(t) \cdot \delta {\bf r}_j(0) \rangle/a^2 $,
\begin{equation}\label{eq:gt2}
\partial_t G(t) = \sigma \widetilde{\gamma}^{-1} \Gamma^{(0)} G(t)
\end{equation}
where 
\begin{equation}\label{eq:sigma}
\sigma = \frac{3 k_B T}{a^2 \gamma_0} = \frac{3 D_0}{a^2}.
\end{equation}
sets the time scale for the relaxation.

Since Eq.(\ref{eq:gt2}) is linear, it can be solved by transforming 
to normal modes by a similarity transform: 
\begin{equation}\label{eq:lambda}
Q^{-1}[\tilde{\gamma}^{-1} \Gamma^{{\rm (0)}}]Q 
= {\rm diag}\{\lambda_p\}, 
\end{equation}
where the columns of $Q$ are the right eigenvectors of 
$[\tilde{\gamma}^{-1} \Gamma^{{\rm (0)}}]$.  
Even though $\tilde{\gamma}$ and $\Gamma^{{\rm (0)}}$ are 
symmetric, their product
is not symmetric in general.
Nevertheless, $Q$ can diagonalize
$\tilde{\gamma}$ and $\Gamma^{{\rm (0)}}$ separately (though not by 
a similarity transformation
since $Q$ is not orthogonal in general) \cite{yamakawa:71}. With 
\begin{equation}\label{eq:eta_nu}
Q^T \tilde{\gamma} Q = {\rm diag}\{\nu_p\}
\; \; \; \; \;
Q^T \Gamma^{{\rm (0)}} Q = {\rm diag}\{\eta_p\}
\end{equation}
we have the relationship
\begin{equation}\label{eq:moderel}
\lambda_p = \eta_p/\nu_p.
\end{equation}
The correlations between monomers, expanded onto the normal modes is
\begin{equation}\label{eq:gt_modes}
G_{ij}(t) 
= \sum_{p} Q_{pi}Q^T_{jp}/\eta_p \exp (- \sigma \lambda_p t).
\end{equation}
We note that equal time correlation functions are the equilibrium
correlations $G_{ij}(t=0) = G_{ij} \equiv [\Gamma^{{\rm(0)}}]^{-1}$.

The reference Hamiltonian has two features that distinguish this chain 
model from the nearest neighbor Rouse chain: chain stiffness, 
parameterized by $g$, and constraints to the native structure, specified
by $\{C\}$.  
We consider the effect on the pair correlations 
of these two contributions separately, to give a point of 
reference for the model. To simplify this illustration of the dynamics, 
we consider a (free-draining) diagonal friction matrix,  
$\tilde{\gamma}_{ij} = \delta_{ij}$.  Then, 
$\nu_p = 1$ which implies that the relaxation rates are $\lambda_p = \eta_p$
and the normal modes are the eigenvectors of the chain connectivity 
$\Gamma^{{\rm (ch)}}$.

For an unconstrained polymer ($C_i = 0$), the dynamics of the stiff, 
freely rotating chain has been studied quiet extensively 
\cite{mb:rz:78,ga:fg:81,ap:sb:cc:84,ap:mg:86,ap:fg:ga:87}.
Referring to the definition of $\Gamma$ 
[Eq.(I-4)], 
$Q$ differs from that for the well known rouse 
modes $Q^R$ 
\begin{equation}\label{eq:rousemodes}
[Q^R]^T K^R Q^R = {\rm diag}\{ \eta^R_p \}
\end{equation}
because of the boundary term, $g^2/(1-g)\Delta$, which is only important 
for very stiff chains \cite{ap:sb:cc:84}.  Thus, unless 
$g \approx 1$, $Q \approx Q^R$ for the free-draining,
unconstrained polymer:
\begin{equation}\label{eq:r_modes}
Q^{{\rm R}}_{pi} \sim  \cos \left[ \frac{\pi p}{n} (i - 1/2) \right]
\hspace{1cm}
\eta^{{\rm R}}_p = 4 \sin^2 \left[ \frac{\pi p}{2n} \right]
\end{equation}
where the mode index takes values $p = (0,\ldots,n-1)$. 
The relaxation spectrum is then approximately
\begin{equation}\label{eq:c0lambda}
\lambda_p = \eta_p \approx \frac{1-g}{1+g} \eta_p^R
+ \frac{g}{1-g^2}(\eta_p^R)^2 + B
\end{equation}
where the final term is due to the confinement term in 
Eq.(I-9). 
The second term of Eq.(\ref{eq:c0lambda}) causes
the short wavelength modes (high mode index $p$) to relax faster 
as the chain stiffness increases.  In contrast, the relaxation rates for the 
long wavelength modes (low mode index $p$) rates are suppressed by increasing 
chain stiffness.  The number of modes with slowly relaxing rates
that are suppressed
diminishes as the chain stiffness increases; for the limiting stiffness
corresponding to a rigid rod, $g \rightarrow 1$, the relaxation rates for
all modes except one diverge \cite{mb:rz:78}.  

\begin{figure}
\psfig{file=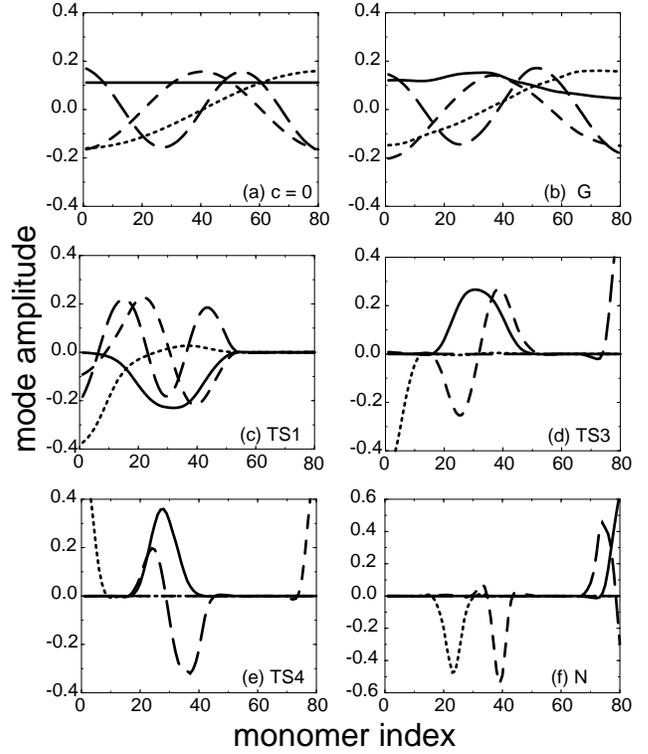,width=3.3in}
\caption
[Slowest modes for constrained ensembles.]
{
Mode amplitude plotted vs. sequence index for the four modes with
the slowest relaxation rates; 
from the slowest mode: solid, dashed, long dashed, dotted. 
($\gamma_I = 0$.)
The constraints correspond to stationary points
on the free energy surface for $\lambda$-repressor with persistence
length $l = 5\,a$ ($g = 0.8$): (a) unconstrained,(b) Globule, (c) $TS_1$,
(d) $TS_3$, (e) $TS_4$, (f) Native. 
}
\label{fig:slomodes}
\end{figure}

When the polymer is 
constrained by a non-uniform pattern of nonzero constraints 
$\{C_i\}$,
the normal modes are not the Rouse modes of a free chain. 
Fig.\ref{fig:slomodes} 
shows the four slowest modes for a chain with stiffness parameter
$g = 0.8$ (persistence length $l = 5\, a$) and various constraints.
In addition to the modes for the unconstrained chain shown in 
Fig.\ref{fig:slomodes}a, 
the modes corresponding to selected stationary points along
the folding route of $\lambda$-repressor 
are also shown in Fig.\ref{fig:slomodes}b--f.  
The slowest modes for the denatured state 
(Fig.\ref{fig:slomodes}b)
differ only slightly from those for the unconstrained
chain (Fig.\ref{fig:slomodes}a) given by Eq.(\ref{eq:r_modes}).
Referring to the fluctuations shown in 
Fig.I-9b,
the structure at the transition state $TS_1$ can be roughly
described by saying that residues in helices H4-H5 (indices 53--80) are
localized, while the residues in helices H2--H3 (indices 27--47)
have the largest fluctuations%
\cite{fn:ptw00b_lam}
.
These constraints are reflected in
the structure of the slowest relaxation modes show in 
Fig.\ref{fig:slomodes}c. The motion of the more localized residues is
shifted to modes with higher relaxation rates so that
these residues are absent in the slowest relaxation modes.
In addition, the mode with the slowest relaxation (solid line) 
is no longer a uniform translation mode, but involves motion of 
the most unconstrained resides 
in helices H2--H3 (indices 27--47). The other modes are much the 
same as those
of an unconstrained but shorter chain with one end fixed
at the end of helix H4. 
Similar correspondence between the mode
structure and the localization of residues in the transition state
ensemble can be seen for the 
constraints corresponding to $TS_3$ and $TS_4$
in Fig.\ref{fig:slomodes}d--e where the modes become increasingly 
more localized in sequence due to non-sequential constrained
residues.
The normal modes for the native state (Fig.\ref{fig:slomodes}f)
are more localized with the slowest relaxation involving
localized regions with the largest temperature factors.

\begin{figure}
\psfig{file=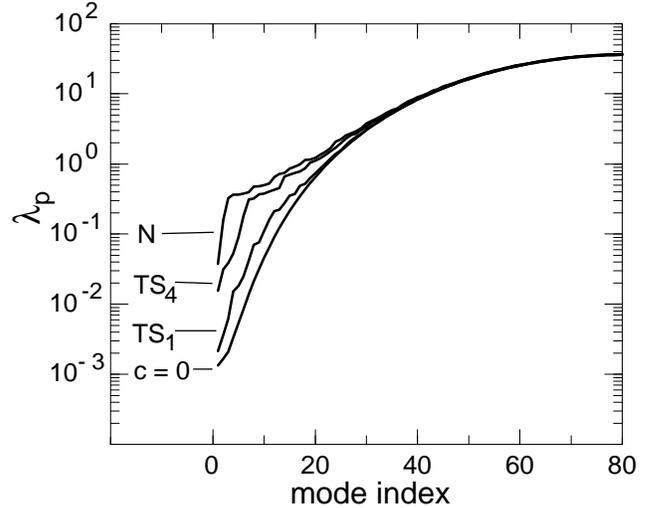,width=3.3in}
\caption
[Monomer relaxation rates for constrained ensembles.]
{
Relaxation rates $\lambda_p$ vs. mode index for different constraints
corresponding to stationary points
on the free energy surface for $\lambda$-repressor with persistence
length $l = 5\,a$ ($g = 0.8$): unconstrained ($C = 0$), 
$TS_1$, $TS_4$, and Native
(N) are indicated in the plot.
}
\label{fig:tinv.gst.8}
\end{figure}

The mode spectra (relaxation rates) for the free chain, constraints
corresponding to transition states, and the
native state are shown in Fig.\ref{fig:tinv.gst.8}.  
As can be seen in this figure, while the spectrum for the
short wavelength modes (large mode index) are relatively unaffected,
the relaxation rates for all the long 
wavelength modes (small mode index)
become faster as the protein takes on more structure, 
reflecting the relatively rapid motion toward the  
native structure as the constraints provided by the native
structure become stronger.

\subsection{Internal friction}

One can expect that
Eq.(\ref{eq:gt2}) with harmonic forces can capture the correct
physics of the low frequency spectrum of the chain dynamics, but, 
as in the Debye theory of crystals, the description will not be correct 
for small length scales which depend on microscopic details. 
For many polymer physics applications, even though the polymer has a
broad range of relaxation rates, the experimentally observed dynamics 
is dominated by the 
the long wavelengths (in sequence) motion corresponding to the slowest modes.  
de Gennes calls the independence of the long wavelength modes on the 
microscopic conformational changes the ``Kuhn Theorem'' \cite{deGennes:79}. 
Why then is it interesting to incorporate internal viscosity (arising from local
activated isomerization) into a model of
barrier crossing dynamics?  Because we do not know a priori the
scale of the motions involved in locally crossing the transition
region.

To be more specific,
we consider internal friction for the following two reasons: 
(1) We have seen that the nonuniform constraints
to the native structure affects the structure of the normal modes 
moving them to involve more local motions.
It is unclear then how the Kuhn theorem (which is a statement about
the delocalized continuum limit) applies here.
(2)  We are ultimately interested in studying the 
barrier crossing rate for protein folding, through the Laplace transform 
of the native density correlations [Eq.(\ref{eq:rhocorr})]
which mix different Rouse modes. 
A correlation function similar to the the native density
correlation function arises (for an unconstrained polymer model)
in the Wilemski-Fixman theory of diffusion-limited chain closure reactions
\cite{gw:mf:73a,gw:mf:73b}. 
In analyzing this theory, Doi showed that for end-to-end monomers,
the Laplace transform evaluated at small frequencies is actually dominated by the 
integral at short times,
i.e., all modes contribute to the integral, not just those with the
longest relaxation times \cite{doi:75e}.

The friction matrix in Eq.(\ref{eq:gt2}) 
$\gamma = \gamma_0 \widetilde{\gamma}$ is written a sum of two contributions
\begin{equation}
\gamma = \gamma^{{\rm S}} + \gamma^{{\rm IV}}
\end{equation}
where the first term is proportional due to the solvent viscosity, and the 
second term accounts for the ``internal viscosity'' and is taken to be 
independent of the solvent friction.  
A typical choice for the effective solvent friction 
is the Oseen tensor from the hydrodynamic mobility \cite{doi_edwards:86}.
This non-local friction mediated by the solvent damps the motion of 
one bead due to the motion of another distant beads. It is not difficult to 
use standard preaveraging approximations to incorporate these 
hydrodynamic interactions.
For simplicity, we neglect this aspect of the 
problem and consider the free-draining approximation
\begin{equation}
\gamma_{ij}^{{\rm S}} = \gamma_0 \delta_{ij}
\end{equation}
where $\gamma_0$ is defined in Eq.(\ref{eq:gamma_0}).

As previously discussed, some internal friction is due to 
the neglect of the activated character of the individual conformational
transitions of the backbone angles.
While there are other ways to incorporate this time scale into a model of
chain dynamics \cite{allegra:86}, 
perhaps the most simple correction to the Gaussian description is through
the friction matrix.  In accordance with the Kuhn theorem, the long wavelength 
motion should have the usual solvent viscosity dependence being largely 
unaffected by the microscopic barrier crossing; on the other hand, short 
wavelengths should tend to relax more slowly, due to the ``dynamical stiffness'' 
caused by the slow activation rates for dihedral angle changes.
Since the dihedral angle potentials are quite steep these will have a much
weaker viscosity dependence.

In the model presented in Ref. \citen{erb:mcw:73}, 
Baz\'{u}a and Williams derived an effective frictional force on the 
$i^{\rm th}$ bead which acts on the relative velocities of adjacent beads:
\begin{equation}\label{eq:fiv0}
F_{{\rm IV}} =
\gamma_I ( \dot{{\bf r}}_{i+1} - \dot{{\bf r}}_{i})^{{\rm d}}
- \gamma_I ( \dot{{\bf r}}_{i} - \dot{{\bf r}}_{i-1})^{{\rm d}},
\end{equation}
where the superscript denotes the projection of the forces along the
bond directions.  A common approximation is to write the projected
velocities as a sum of the difference between the full velocity vectors 
$\{\dot{{\bf r}}\}$ and the difference between 
angular velocity of the beads \cite{cwm:mcw:89}.  
The angular component is important for  inherently anisotropic 
phenomena such as flow birefringence, but for our purposes we can 
neglect the angular rotation component and write the internal friction 
in the form \cite{cwm:mcw:85,fixman:87,fixman:89,erb:mcw:73}
\begin{equation}
F_{{\rm IV}} = - \gamma_I \sum K^{R}_{ij} \partial_t(\delta {\bf r}_j),
\end{equation}
where $K^R$ denotes the Rouse connectivity matrix 
[Eq.(I-5)], 
and we have used $\partial_t({\delta \bf r}_i) = \partial_t {\bf r}_i$  
since the average monomer position is constant for fixed constraints.

The total friction matrix is the sum of the 
(free-draining) diagonal solvent friction and the internal friction
\begin{eqnarray}\label{eq:gammai}
\gamma_{ij} & = & \gamma_0 \delta_{ij} + \gamma_I K^{{\rm R}}_{ij} \nonumber \\
            & = & \gamma_0 (\delta_{ij} + \gamma_I/\gamma_0 K^{{\rm R}}_{ij})
\end{eqnarray}
which defines the dimensionless friction matrix $\tilde{\gamma}$ in 
Eq.(\ref{eq:gamma_0}).

The spectrum of relaxation rates with nonzero internal friction 
can be illustrated by considering an unconstrained polymer chain ($C_i = 0$).
From Eq.(\ref{eq:moderel})
the mode relaxation rates for this case is approximately
\begin{equation}\label{eq:modeiv}
\lambda_p \approx  \frac{\eta_p}{1 + \gamma_I/\gamma_0 \, \eta_p^R},
\end{equation}
where $\eta$ are the eigenvalues of $\Gamma^{(0)}$ 
(Eq.(\ref{eq:c0lambda}))and $\eta_p^R$ are the eigenvalues of the Rouse 
matrix $K^R$. This equation shows that the internal friction leaves the 
rates of the long wavelength modes (small $\eta^R$) relatively unchanged;
this is the Kuhn theorem.
The short wavelength modes (larger $\eta^R$), are suppressed by
a non-vanishing internal friction.
The crossover between the two behaviors,
hydrodynamic solvent limited versus conformational interconversion limited,
is determined by the value 
of $\gamma_I/\gamma_0$.  

\begin{figure}
\psfig{file=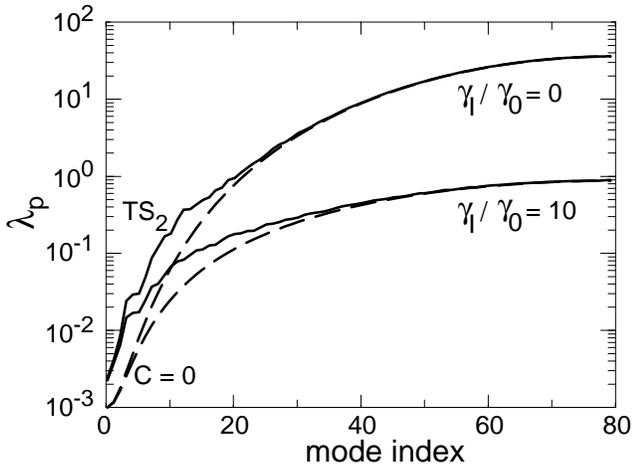,width=3.3in}
\caption
[Monomer relaxation rates for different values of internal viscosity]
{
Relaxation rates $\lambda_p$ vs. mode index for two values of internal
viscosity: $\gamma_I/\gamma0 = 0$ and $\gamma_I/\gamma0 = 10$ (indicated 
on the plot).
The constraints correspond to $\lambda$-repressor with persistence
length $l = 5\,a$ ($g = 0.8$): unconstrained (dashed lines) and 
$TS_2$ (solid lines).
}
\label{fig:tinv.gst.8.iv}
\end{figure}

For constrained chains, the effect of internal friction 
on the relaxation modes is different.  The internal friction
damps rapidly oscillating features of the modes which even
occur in the relaxation modes with long relaxation times
since they are not identical to Rouse modes.
(Note the  modes illustrated in Fig.\ref{fig:slomodes}, 
which are also different from Rouse mode were calculated 
for $\gamma_I = 0$).  
Nevertheless, for the constraints considered here,
this seems to be a minor effect, and the relaxation modes still
follow the general behavior of Eq.(\ref{eq:modeiv}).
Fig.\ref{fig:tinv.gst.8.iv} shows the relaxation spectrum with
zero and finite internal friction
for the unconstrained chain and for the constraints corresponding to the 
transition state $TS_2$ of the variational calculation for 
$\lambda$-repressor. 
The spectrum corresponding to the other transition states
is similar: the relaxation rates for the slow modes are relatively
unaffected, whereas the faster modes have suppressed relaxation rates.

\section{Folding Prefactor Example: $\lambda$-repressor}
The average folding route 
for the fast folding variant of the $\lambda$-repressor protein presented in [I]
is characterized by sets of constraints $\{C\}$
corresponding to local minima and saddle-points of the variational free energy.
For each saddle-point,  the pair correlations $G(t)$ are
calculated through the normal mode expansion given by  Eq.(\ref{eq:gt_modes}) 
with the friction matrix given by Eq.(\ref{eq:gammai}).  
These monomer correlations determine the native density correlations
${\cal C}(t)$ through Eq.(\ref{eq:drhocorr}) 
[with Eq.(\ref{eq:rhoave}) and Eq.(\ref{eq:rhocorr})], and its Laplace transform 
$\widehat{{\cal C}}(\omega)$ is taken numerically.
With the mobility matrix $\widehat{\mu}(\omega)$ given by Eq.(\ref{eq:mu_rho}) and 
the curvature matrix of the free energy $\bar{\Gamma}^{\star}$ given by
Eq.(\ref{eq:chainrule}),  the eigenvalue equation Eq.(\ref{eq:usmode}) 
is then solved numerically giving the reaction coordinate and folding rate 
prefactor for that transition state.

\begin{figure}
\psfig{file=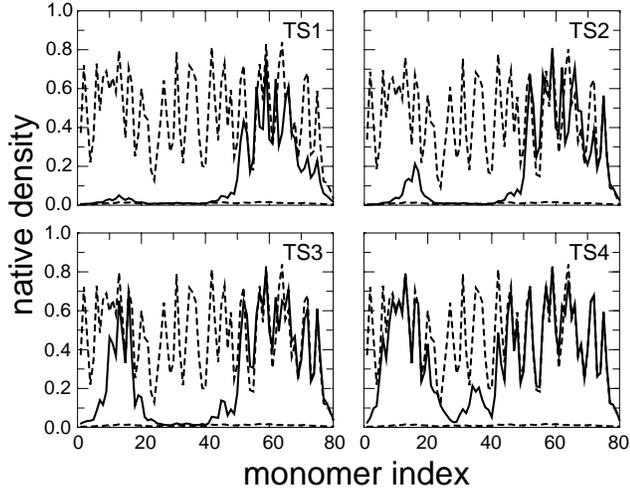,width=3.3in}
\caption
[Native density of constrained ensembles.]
{
Native density $\rho_i$ evaluated at the native and globule 
minima (dashed) and the transition states $TS_1$--$TS_4$ (solid) 
of the average folding route for $\lambda$-repressor.  
The persistence length of the chain
is $l = 5 \, a$ ($g = 0.8$) and the width defining 
$\rho_i$ is set by $\alpha = 0.5$.
}
\label{fig:natdens}
\end{figure}

In this example, 
we consider the folding routes obtain for a chain with persistence 
length of polyalanine, $l = 5 \, a$ ($g = 0.8$).
The native density $\{\rho\}$ also depends on the parameter 
$\alpha^{\mathrm{N}}$ controlling the width of the Gaussian measure to the native 
structure.  This width should be small enough to characterize the
native structure accurately, but the appropriate value is also limited by 
the resolution of the model 
(approximately equal to the root mean square (RMS) fluctuations of the bond
length $a$).
For large $\alpha^{\mathrm{N}}$ (small width), 
$\{\rho\}$ has low values even for the
native state since the average positions are not precisely equal to the native
structure at finite temperature; for $\alpha^{\mathrm{N}} = 1$ 
the average value of $\rho_i$ evaluated at the
native state is 0.3.
For the native densities shown in Fig.\ref{fig:natdens} with
$\alpha^{\mathrm{N}} = 0.5$, the average of $\rho_i$ at the
native state increases to $\rho_i$ is 0.5.  
We consider this to be a reasonable value
to use for $\alpha^{\mathrm{N}}$.
For smaller $\alpha^{\mathrm{N}}$ (larger widths), the separation
between
$\{\rho\}$ evaluated at the native and globule state is more distinct,
though unfolded regions of the transition states have native densities
closer to the native values.  
The dependence of the prefactor on the width $\alpha^{\mathrm{N}}$ 
is considered below.

\begin{figure}
\psfig{file=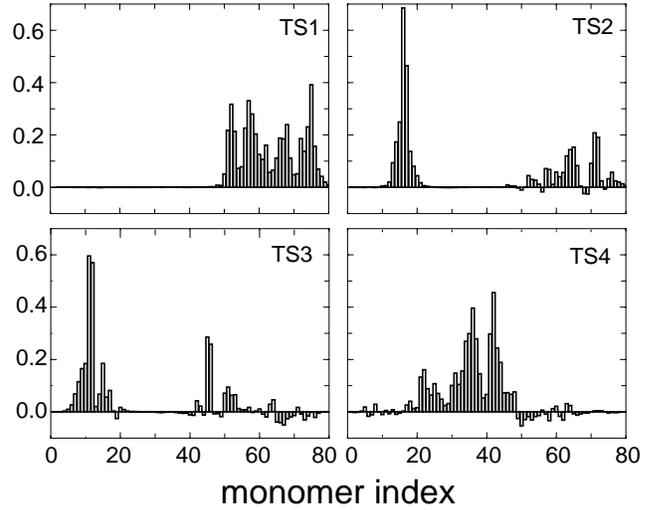,width=3.3in}
\caption
[Local reaction coordinate for transition state ensembles.]
{
Local reaction coordinates for the transition states $TS_1$--$TS_4$ 
of the $\lambda$-repressor average folding route at $T_{\mathrm{f}}$.  Normalization of
the reaction coordinate is set to $\sum {\bf u}_i^2 = 1$.
Parameters are the same as in Fig.\ref{fig:natdens}.
}
\label{fig:usmodes}
\end{figure}

The local reaction coordinates (unstable modes) for the four transition 
states along the folding route are shown in Fig.\ref{fig:usmodes}. 
These unstable modes are quite understandable in terms of the structure inferred
by the temperature factors in 
Fig.I-9b.
The unstable mode of $TS_1$ (Fig.\ref{fig:usmodes}a)
is rather delocalized, consisting of residues in helices H4--H5 and modulates with
a wavelength approximately equal to the persistence length $l/a = 5$; the
partial localization of helix H1 indicated by the temperature factors is not a
part of the normal mode.  The mode for $TS_2$ (Fig.\ref{fig:usmodes}b) 
has a significant component of formation of helices H4--H5 as in $TS_1$, 
but the localized 
region around a residue in H1 is a much larger component.  The unstable mode at
$TS_3$ is also localized, but consists mainly of a few residues separated
in sequence in helix H1 and near helix H4. 
Finally, the unstable mode at $TS_4$ is delocalized 
consisting of residues in helices H2--H3 which are the last helices to
form.
The features of the reaction coordinates resemble the unstable modes of 
the free energy ($\bar{\Gamma}$), though the relative amplitudes of the 
peaks in a given reaction coordinate are altered by the chain dynamics through
the mobility $\widehat{\mu}(\omega)$.

Since the local reaction coordinates along the 
average folding route are very different, one from each other, 
a reduced description of the kinetics based on a single reaction coordinate
can not be used when dynamics is examined at this level of detail.  Attempts
to find perfect global dynamical coordinates in lattice simulations of perfectly 
funneled models are consistent with this \cite{vsp:dsr:99,dpgts:98}.
In the present description, each dynamical reaction coordinate applies
in the vicinity of one saddle-point.
The changes of the local reaction coordinate along the folding route
are intimately related to the fine structure of the free energy profile.
Notice that the effects on predicted reaction rates are minor, however.
Surmounting the highest barrier gives the rate within a factor of two
(vide supra).
Calculations of one-dimensional barrier crossing with shallow local minima
(intermediates) also show why having a perfectly good 
local detailed description of the free energy profile need not yield a strictly
defined global reaction coordinate \cite{cw:tk:99}.
At the same time, a much cruder reaction coordinate can function
perfectly well for predicting rates if the appropriate configurational
diffusion coefficient is used \cite{nds:jno:pgw:96}.

\begin{figure}
\psfig{file=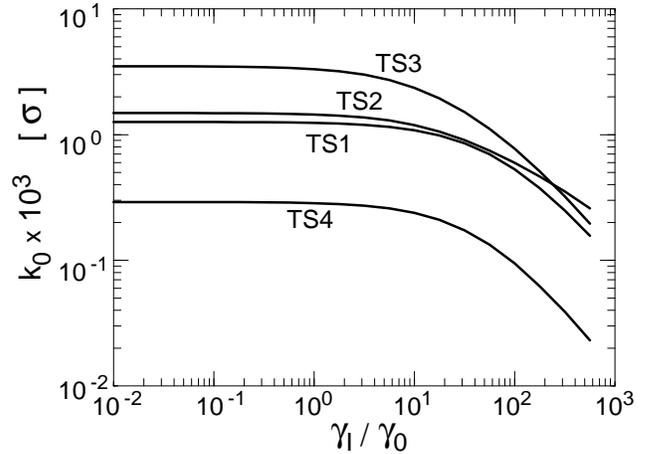,width=3.3in}
\caption
[Prefactors as a function of internal friction.]
{
Prefactor $k_0 = |\kappa^{\star}|/2\pi$ in units of $\sigma = 3 D_0/a^2$
as a function of 
$\gamma_I/\gamma_0$ for the transition states $TS_1$--$TS_4$ 
of the $\lambda$-repressor average folding route at $T_{\mathrm{f}}$.
Parameters are the same as in Fig.\ref{fig:natdens}.
}
\label{fig:k0_vs_dynst}
\end{figure}

The prefactor $k_0 = |\kappa^{\star}|/2\pi$ for each transition state is plotted
as a function of the internal friction $\gamma_I/\gamma_0$ in 
Fig.\ref{fig:k0_vs_dynst}.
When the unstable curvature of the free energy is
large, there is less re-crossing at the top of the barrier leading 
to a larger prefactor.  
For small $\gamma_I/\gamma_0$ the prefactors increase in the same
order as the corresponding curvatures of the free energy ($\bar{\Gamma}_0$):
$k_0^{(4)}< k_0^{(1)}<k_0^{(2)}<k_0^{(3)}$.  The relative values of 
the prefactors at different transition states
depend on the chain dynamics through the mobility matrix
$\widehat{\mu}(\omega)$, however, and do not follow the
proportions of the corresponding free energy curvatures.
For small values of $\gamma_I/\gamma_0$, the prefactor is 
proportional to the bare diffusion coefficient $D_0 = k_B T/\gamma_0$.
The prefactors decrease with increasing $\gamma_I/\gamma_0$ because
of the suppressed relaxation rates of the fastest modes.  
This turnover occurs around $\gamma_I/\gamma_0 = O(1)$, though it 
is different for each transition state.

To put these results in laboratory units for polypeptides, 
we need to specify the monomer diffusion coefficient 
$D_0$ and the internal viscosity parameter
$\gamma_I/\gamma_0$.  We estimate the typical hydrodynamic radius
in Eq.(\ref{eq:gamma_0}) to be the sum of the monomer spacing $a = 3.8$\AA\
and the average van der Waals radius of a typical side chain
$a_{\mathrm{vdw}} \approx 3$ \AA\ \cite{creighton:93}: 
$a_{\mathrm{eff}} \approx 7$\AA.
Using the viscosity of water at room temperature
$\eta_0 = 1$ centipoise gives the a monomer diffusion coefficient 
$D_0^{\rm H_2O} = 3 \times 10^{-6} \mathrm{cm}^2/\mathrm{s}$.
We estimate the internal viscosity by the elementary bond flipping rate 
of the polymer backbone.  Calculations of the 
$\beta \rightarrow \alpha$ 
conformations of di-alanine give a value for 
$\tau_{\mathrm{flip}}^{-1} \approx 10 \mathrm{ns}^{-1}$ \cite{pgb:cd:dc:00}.  
One can expect this rate to be reduced when the flipping transition
occurs in a long polypeptide
chain since the neighboring monomers must adjust to some extent in order
to minimize the motion of the chain through the solvent \cite{js:eh:80}.
Judging from studies on model chains one expects about a factor 
of three slower,
we estimate a typical value to be 
cite{js:eh:80,eh:js:82}
$\tau_{\mathrm{flip}}^{-1} = 3 \, \mathrm{ns}^{-1}$ 
. 
It would be nice for simulators to pin this number down for the peptide
isomerization rate in long chains.
It is obvious that the relaxation rates describing the chain dynamics should
not exceed this elementary rate.  Associating the internal friction with
this isomerization rate, we set $\gamma_I/\gamma_0$ so that the free chain
relaxation rates of the fastest mode equals $\tau_{\mathrm{flip}}^{-1}$. 
For a chain with persistence length
$l = 5 a$ and monomer diffusion coefficient $D_0^{H_2O}$, we find $\gamma_I/\gamma_0 = 9.1$.

\begin{figure}
\psfig{file=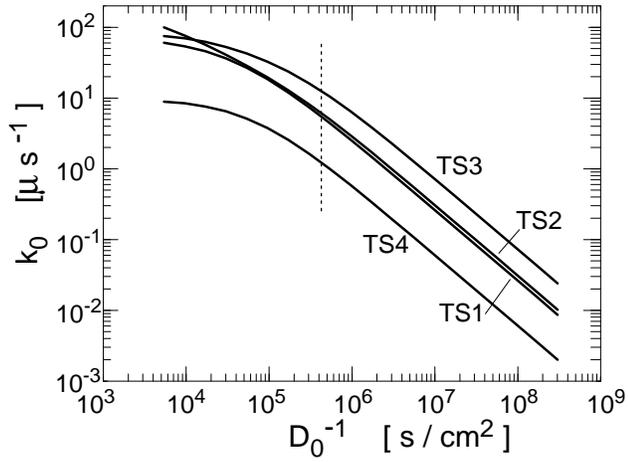,width=3.3in}
\caption
[Prefactors as a function of the inverse diffusion coefficient.]
{
Prefactor $k_0$ is plotted as a function of the inverse
monomer diffusion
coefficient $D_0^{-1} = \gamma_0/k_B T_{\mathrm{f}}$.
The internal friction is set equal to the fixed 
value $\gamma_I/k_B T_{\mathrm{f}} = 9.1/D_0^{\mathrm{H_2O}}$.
The vertical dotted line corresponds 
$D_0 = D_0^{\mathrm{H_2O}} = 3 \times 10^{-6} \, \mathrm{cm}^2/\mathrm{s}$
(and $\gamma_I/\gamma_0 = 9.1$).
Parameters are the same as in Fig.\ref{fig:natdens}.
}
\label{fig:k0_vs_D0inv}
\end{figure}

In Fig.\ref{fig:k0_vs_D0inv}, the prefactor $k_0$ is plotted against
the inverse monomer diffusion coefficient $D_0^{-1}$ in laboratory units.  
The internal friction is set to the fixed value $\gamma_I = 9.1 \gamma_0$ 
for $\gamma_0/k_B T_{\mathrm{f}} = (D_0^{\mathrm H_20})^{-1}$.  The prefactor is 
proportional to the solvent viscosity ($D_0^{-1}\sim \eta_0$) for large
viscosities and saturates at low viscosities.  
This turnover is reminiscent of the viscosity dependence of the 
prefactor from Kramer's theory \cite{kramers:40}, 
but here the turnover is driven indirectly by the internal 
friction of the chain
rather than explicit inertial terms reflecting an energy limited regime. 
The prefactor in pure water,
indicated by the dashed line in Fig.\ref{fig:k0_vs_D0inv}, is still in
the linear regime, but  near the turnover.
The prefactors of the different transition states for $D_0 = D_0^{\mathrm{H_2O}}$
(given in Table \ref{tab:k0_tf})
are on the order of $10 \mu \mathrm{s}^{-1}$.
As a point of reference, the unstable mode frequency for $TS_2$, 
$|\kappa^{\star}| = 6.9 \times 10^{-3} \, \sigma$, 
is approximately equal to the relaxation rate of the $p = 6$
normal mode of a free chain of $n = 80$ monomers and persistence
length $l = 5 \, a$
($\lambda_{p = 5} = 7.2 \times 10^{-3}$). 
The normal mode for this motion has a 
node to node period of 20 monomers, twice the length used in
the estimates based on the smallest loop closure \cite{zg:dt:95,hhse:96}.

The folding rate includes the activation barrier (calculated with this model
in [I]) as well as the prefactor.
The calculated thermodynamic and kinetic quantities needed for the folding rate
are given in Table \ref{tab:k0_stab} for two thermodynamic conditions.
Adjusting the energy scale of the contacts $\epsilon_0/k_B T$ fixes the
equilibrium constant $K = k_f/k_u$ (the ratio of the folding and unfolding rates),
so that $\ln K$ gives the difference in free energy between the
globule and native state (in units of $k_B T$).  
For the very stabilizing condition
$\ln K = 7.8$, the prefactor is $k_0 = 2.4 \mu \mathrm{s}^{-1}$ and the 
barrier is $\Delta F^{\dagger}/k_B T = 3.4$, giving a folding time of
$1/k = 13 \mu \mathrm{s}$.  This agrees well with the experimental rate 
$1/k = 12 \mu \mathrm{s}$ at this stability 
(extrapolated to 0M denaturant) \cite{bhdfo:96}.
At the transition midpoint $\ln K = 0$, the calculated prefactor 
is $7.6 \mu \mathrm{s}$, and with the barrier height 
$\Delta F^{\dagger}/k_B T = 5.1$, we get a folding time 
of $1/k = 23 \mu \mathrm{s}$  which is considerably faster than the measured
folding time of $1/k \approx 10 m \mathrm{s}$ \cite{bhdfo:96}.
This suggest the microscopic calculation for the perfectly funneled
surface under estimates the barrier by about $5 k_B T$ at the folding
transition temperature.

The calculated inverse prefactor for $\lambda$-repressor
at the transition midpoint 
$1/k_0 \approx 0.1 \mu \mathrm{s}$ and under highly stabilizing 
condition  $1/k_0 \approx 0.5 \mu \mathrm{s}$ are both
faster than the estimated fastest possible folding time  
\cite{zg:dt:95,hhse:96}
$\tau \approx 1 \mu \mathrm{s}$.
The latter is the time it takes to form a typical contact 
as determined by the fluorescence or chemical 
quenching rate of a pair of residues
separated by a typical loop length.
It is very difficult to imagine that folding
can occur on times faster than contact formation, and the 
introduction of this speed limit is important conceptually
as a reference timescale for protein folding dynamics.
It is tempting to identify the speed limit with the folding prefactor 
since the fastest rate is obtained from Eq.(\ref{eq:rate}) for
a vanishing free energy barrier.
But this is not precisely correct since the separation of time
scales assumed in activated kinetics is not true for small barriers.
As seen in Table \ref{tab:k0_stab}, reducing the activation barrier
(by stabilizing the native state, for example) changes
the prefactor since the saddle--point curvature 
of the free energy changes.
If the prefactor at both stabilities in this example
were $1/k_0 = 0.1\mu \mathrm{s}$,
the folding time with the reduced activation barrier 
would be a factor of approximately 6 times smaller than at 
the transition midpoint,
or  $1/k \approx 4 \, \mu \mathrm{s}$, which approaches the speed limit.
The prefactor is found to be approximately three times smaller 
than at the midpoint, however, resulting in a larger folding time 
$1/k \approx  12 \, \mu \mathrm{s}$.
Pushing the conditions to the limit of a vanishing barrier, 
it could happen that the barrier is very low but still has a large 
unstable mode free energy curvature, resulting in a 
calculated folding time somewhat faster
than the speed limit. For such near downhill folding \cite{bosw:95,js:je:mg:99}, 
however, the folding would not be single exponential process dominated by 
barrier crossing, but involves kinetics more directly 
connected to the time for contact formation than
the time reflected in the apparent prefactor.

It is interesting to consider
the dependence of the prefactor on the chain dynamical parameters
since the values for the monomer diffusion coefficient and 
the isomerization rate in a polymer chain are not known precisely 
(see next section). Conservatively, the values we used are probably
accurate to within a factor of two or three. 
As long as the internal friction is not too large,
the monomer diffusion coefficient is the more important parameter for 
the absolute prefactor (in $\mathrm{s}^{-1}$);
since the time scale for the monomer relaxation is 
$3 D_0/a^2$, the absolute prefactor is proportional to this scaling.
Table \ref{tab:k0_vary} gives the prefactor from $TS_2$ using
the free alanine diffusion coefficient and the di-alanine isomerization rate.
As shown in Table \ref{tab:k0_vary}, the prefactor varies by about
a factor of three when $D_0^{\mathrm{H_2O}}$ is the diffusion coefficient 
of a free alanine molecule.
This is the same factor as the ratio of the monomer diffusion coefficient,
though it also depends 
on the value of the internal friction (i.e., $\tau^{-1}_{\mathrm{flip}}$)
as well.

\begin{figure}
\psfig{file=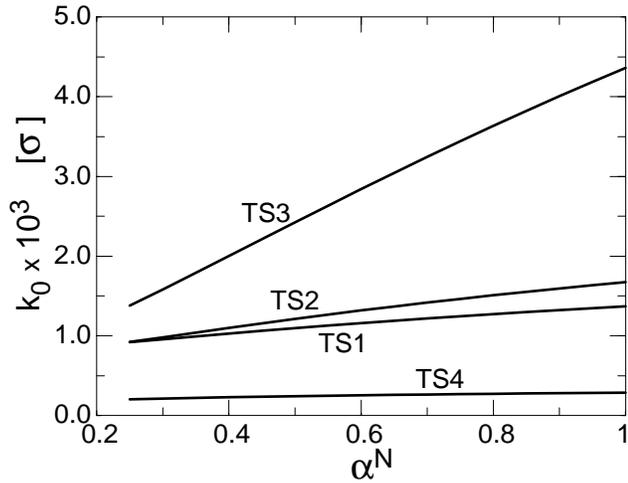,width=3.3in}
\caption
[Prefactors as a function of the Gaussian width of the native density.]
{
Prefactor $k_0 = |\kappa^{\star}|/2\pi$ vs. the Gaussian width of the native
density $\alpha^{\mathrm{N}}$.  The persistence length of the chain is $l = 5\,a$ 
($g = 0.8$) and the internal friction is $\gamma_I/\gamma_0 = 0.91$.
}
\label{fig:k0_vs_alpha}
\end{figure}

The value of the Gaussian width 
$\alpha^{\mathrm{N}}$ defining the native density is 
a parameter of the model, not a measurable quantity, so we must also
consider the sensitivity of our results to the value of $\alpha^{\mathrm{N}}$.
As shown in Fig.\ref{fig:k0_vs_alpha},
the prefactors $k_0$ are approximately linear in $\alpha^{\mathrm{N}}$, 
increasing by a factor of about two for $TS_3$ and less than a 
factor of unity for the other transition states
as $\alpha^{\mathrm{N}}$ varies from 0.25 to 1.0.
The different slopes are reflect the different increases in curvatures
of the free energies as the Gaussian width narrows.
This degree of uncertainty seems reasonable, but the different slopes
suggests that the dependence is a delicate issue, depending
on the detailed structure of each transition state.

\section{Concluding Remarks}

In this paper, we derived a microscopic theory for the dynamics of
protein folding barrier crossing
when non-native contacts and trapping effects can be neglected.  
The variational theory presented in [I] characterizes the average
folding routes as connected minima and transition states. These routes are 
specified through variational parameters constraining the protein 
chain to inhomogeneously order about the native structure.  
The quadratic reference Hamiltonian from this theory along with forces from
solvent and internal friction govern the chain dynamics near the saddle-points.
The barrier crossing dynamics is calculated from a 
generalization of both
Grote-Hynes and Langer rate theories to include a frequency dependent friction
in multi-dimensions.  The memory friction itself is approximated by linear
response theory and exploits the harmonic potential of the model.
The theory gives local reaction coordinates and folding rate prefactors
for specific proteins with known native structure.

For the calculated average folding route of $\lambda$-repressor, we find
a folding prefactor $k_0 \approx 10 \, \mu \mathrm{s}^{-1}$ at the transition midpoint
and $k_0 \approx 1\mu \mathrm{s}^{-1}$ under highly stabilizing conditions.  
Using the free energy barriers from this model,
the calculated folding rate agrees well with the measured rate extrapolated to zero
denaturant, but not at the transition midpoint.
Considering other reasonable values for the parameters 
in the model suggests that the calculated prefactor may be off by a factor of 
about three.  These calculations also suggest that the folding rate 
prefactor near water solvent conditions scales linearly with solvent 
viscosity.  This dependence agrees with experimental studies on other two state
proteins \cite{jsbs:97,kwp:db:98,mj:fxs:99} and simulations of 
off-lattice model proteins \cite{dkk:dt:97}.  
Nevertheless, these conditions are near a turnover regime for which the 
internal friction of the 
polymer chain dominates, ultimately causing the prefactor to saturate
upon decreasing solvent viscosity.
Thus folding of polymers with other backbones may be quite different.

The discrepancy between the calculated and measured folding rate at the transition
midpoint $(T_{\mathrm{f}})$ suggests that the model underestimates the activation barrier or
overestimates prefactor for this stability.  
Since the model has many simplifications that could influence both
the thermodynamics and the dynamics,
it is impossible without further investigation
to identify which primarily needs improvement.  It may be combination of both.

One aspect of the present model is that
the native state is quite a bit more flexible than observed in the laboratory,
with RMS fluctuations
on the order of the distance between adjacent monomers $a$
(the spatial resolution of the the harmonic chain).
In comparison, temperature factors
from x-ray crystallography correspond to RMS fluctuation of less than 
1\AA\ $(a/4)$.
A more realistic non-Gaussian polymer backbone would
increase the resolution of the model.
Such a non-Gaussian model could still be treated using the variational
formalism.
Still retaining the Gaussian description of the chain, multi-body forces arising
from neglected degrees of freedom would also 
tend to increase the rigidity of the native state and are a likely
contributor.
Recent off-lattice simulations have
shown that effective multi-body forces sharpen
interfaces between folded and unfolded regions and increase activation
barriers \cite{mpe:pgw:00}.  
As dicussed in [I], contiguous sequence approximations
\cite{ovg:avf:99,ea:db:99,vm:wae:99}
also give inherently sharp interfaces and larger activation barriers even
for pairwise additive forces since the conformations of the residues
are restricted to be either folded or unfolded.
(The contiguous sequence approximations with multi-body forces would give even
higher barriers.)
If multi-body forces are indeed important, the completely folded or unfolded
conformations assumed in
these approximations may well be more appropriate than expected from simulations
of lattice and off-lattice models with only pairwise forces.
Still, the interfaces are probably not as cleanly sharp
as envisioned in the contigous sequence approximations.
Using the calculated prefactor at the transition midpoint, 
$k_0 \approx 8 \mu \mathrm{s}^{-1}$, 
and the measured folding rate of $100 \mathrm{s}^{-1}$ gives
a barrier of approximately $11 k_B T_{\mathrm{f}}$.  This value is greater than both the
activation barrier from the variational 
theory $(\Delta F^{\dagger} \approx 5 k_B T_{\mathrm{f}})$ and 
the double sequence approximation calculated in [I]
$(\Delta F^{\dagger} \approx 8 k_B T_{\mathrm{f}})$.
Multi-body forces may sharpen the distinction between folded and 
unfolded regions in the transition state
ensemble to give an activation barrier consistent with 
the prefactor calculation at the transition midpoint.

Alternatively, it is possible the discrepancy at $T_{\mathrm{f}}$ arises because
of neglected dynamical effects 
that could slow the chain dynamics, reducing the folding rate prefactor.
In this paper, the internal friction 
is interpreted as a correction to account for the activated 
isomerization reaction of the backbone.  As mentioned in the 
introduction, there are other possible contributors to this friction.  
For example, non-native interactions are presently neglected in the
model.  These interactions
contribute to the memory friction and therefore the internal viscosity.
They can be treated through 
mode coupling calculations similar to those we have done for
a random heteropolymer 
\cite{st:jjp:pgw:97} 
(see also Refs. \citen{jrr:eis:96,dt:va:jkb:96,egt:yak:kad:96}). 
A more careful treatment of the solvent may also be necessary for
an accurate description of the chain dynamics.  
Detailed calculations of the reaction dynamics for 
isomerization transition in di-alanine (a much less complicated system) 
suggest the solvent may be an essential component of even the local
reaction coordinate \cite{pgb:cd:dc:00}.  
Furthermore, contact formation dynamics may
also be slowed because contact formation excludes the solvent,
thereby adding an additional activation barrier and 
timescale into the dynamics.  
Supporting this possibility, we note that
simulations of minimal protein folding models with 
desolvation barriers in the
pair potential bring simulations into qualitative agreement with
pressure studies on folding rates \cite{nh:jno:aeg:99}.

Since we expect some of these neglected internal friction contributions to 
influence protein folding dynamics,
we have been careful in this paper to use ``bare'' parameters 
(and investigate other reasonable estimates) instead of treating
the them as fitting parameters in the theory. This is an attempt
to avoid missing a signal that such new dynamical considerations are indeed 
important by using inappropriate effective parameters. 
The folding rate alone, however,
is not enough to  determine the accuracy of the calculated prefactor, since
the barrier height may not yet be accurately enough detetermined.

Consequently, these proposed effects are perhaps best studied at this point
using more direct experimental probes such as 
intrachain fluorescence quenching measurements.  
Fluorescence quenching is a powerful tool to study
the average structure of ensembles \cite{dsg:eh:92,haas:96} 
and has recently been used as
a probe for experiments on single molecules that allow distributions to 
be addressed \cite{jtlldh:99,dlbdmcdsw:00}.
Motivated primarily by the relevance to protein folding dynamics, 
some fluorescence quenching experiments have focused on
less complicated systems such as short peptides 
\cite{eh:ek:izs:78,bwhsdk:99,ljl:wae:jh:99}.
These measurements of simple polymer pair dynamics can guide the
development of improved theoretical models helpful 
to understanding protein folding kinetics.
This is particularly evident in the present model, since
the expression for the effective mobility [Eq.(\ref{eq:mu_rho})] 
is formally related to approximations to the intrachain quenching 
rate in polymers \cite{gw:mf:73a,gw:mf:73b,doi:75e}.

Experiments monitoring the intrachain contact formation through
fluorescence quenching suggest that the dynamical issues mentioned
above play a role in the pair contact dynamics as well.
Fitting data to effective models for the pair dynamics, 
the apparent diffusion coefficient comes out to be 10--20 times smaller
than one would expect from the diffusion coefficients of the 
free probes \cite{hhse:96,dsg:eh:92,haas:96,eh:ek:izs:78,ljl:wae:jh:99}.  
In those models, the polymer is treated as a single pair connected by an
entropic spring controlling the mean square fluctuation of chain ends \cite{ks:zs:as:81}
where the effective diffusion coefficient of the pair, $D_{\mathrm{eff}} = 2 D_0$, 
is a short time approximation to the full
dynamics including all the monomers \cite{rwp:rz:as:96}.
Haas has suggested that the internal friction of the chain
may explain the descrepancy between $D_{\mathrm{eff}}$ and 
the fitted value \cite{eh:ek:izs:78},
but no detailed calculation has been done.  This is a problem
we wish to return to \cite{jjp:pgw:unpublished}.  
Doi's arguments suggest that the short
time properties may dominate \cite{doi:75e}.  We note that
the inclusion of internal viscosity
from the bond flipping rate as presented here $(\gamma_I/\gamma_0 \approx 10)$
reduces the effective short time diffusion coefficient%
\cite{fn:ptw00b_dumbell}
to a value
$D_{\mathrm{eff}} \approx D_0/2$,
narrowing the discrepancy to within a factor of about 2--5.
Since the validity of this short time approximation can be questioned
(e.g., the rate depends on the quenching volume \cite{doi:75e,rwp:rz:as:96}), 
we feel a solution
resolution to this puzzle requires a more complete calculation.

The primary experimental technique used to study the structure
of the transition state ensemble is the $\phi$-value analysis developed by 
Fersht \cite{arf:am:ls:92}.
By comparing the change in the folding rate to the change in stability upon mutation, 
one can infer the extent to which that site is structured in the transition
state ensemble.
To interpret these experiments, one assumes that the mutation alters the
kinetics through a change in the free energy profile, leaving the
prefactor unchanged.  
Sometimes, however, the kinetics and thermodynamics are seen not to be directly coupled.
This may well be a sign of frustration
from non-stabilizing native contacts or non-native contacts influencing the kinetics
of the transition \cite{bas:jw:pgw:99}.
It is also possible however that these mutations affect
the chain dynamics by modifying the internal friction at that residue.
It is easy to extend the present calculation to indicate how  
the internal friction could be exploited to find ``dynamical mutants'' 
in which a residue is replaced by one with higher
dihedral barriers (i.e., a residue with larger local internal friction) perturbing
the prefactor, without altering the stability.
Proline isomerization coupled to folding are an example of this,
but other nonnatural
amino acids may be still more promising probes 
of local reaction coordinates \cite{ngsldl:98}.
Mutations of the n-src loop of the SH3 domain
have very small effect on the equilibrium constant
(preventing the $\phi$-value analysis), but still impact
the folding rate and thus may be examples of dynamical mutants \cite{rgsarb:99}.  
More extreme alterations to the sequence have been 
necessary such as lengthening the loop length in order to characterize 
this region of the transition state 
ensemble \cite{vpg:dsr:db:00}.  
A better understanding of the chain dynamics should help suggest additional 
and more direct probes of similar situations using dynamical mutants.

\section*{Acknowledgments}
This work has been supported by NIH Grant No. PHS 2 R01 GM44557.

\appendix
\section{}
In this appendix, we outline the calculation of the native density
correlations $\langle \rho_i(t) \rho_j(0) \rangle_0$ given by 
Eq.(\ref{eq:rhocorr}).

Since $H_0$ is harmonic, the Green's function
for the diffusion operator is known and has the form of a multi-variable
Gaussian containing all the pair correlations \cite{risken:89}.
The native density correlations
only depend on two monomer positions for a fixed $i$ and $j$, so that 
Eq.(\ref{eq:corrAA_P}) can be simplified by introducing delta functions 
$\int \! \! d{\bf r}_1 d{\bf r}_2 \:\delta({\bf r}_1 - {\bf r}_i)
\delta({\bf r}_2 - {\bf r}_j)$
and integrating over all monomer variables $\{{\bf r}\}$. This reduces
Eq.(\ref{eq:corrAA_P}) to a double integral over ${\bf r}_1$ and 
${\bf r}_2$, and the reduced Green's function contains only
the monomer correlations of the monomer pair, as can be expected
for a Gaussian process.  

It is perhaps easier to make use this Gaussian property from the beginning, 
and obtain the same result as using the full 
Green's function as described above.

Consider a fixed $i$ and $j$ pair.
First we define some notation. We define the following 2-component
vectors:
\begin{equation}
{\bf R} = 
\left[
\begin{array}{c}
{\bf r}_i(t) \\ 
{\bf r}_j(t')
\end{array}
\right]
\hspace{.5cm}
{\bf S} = 
\left[
\begin{array}{c}
{\bf s}_i \\
{\bf s}_j
\end{array}
\right]
\hspace{.5cm}
{\bf R}^{{\rm N}} = 
\left[
\begin{array}{c}
{\bf r}^{{\rm N}}_i \\
{\bf r}^{{\rm N}}_j
\end{array}
\right]
\end{equation}
where, $t$ and $t'$ denote two times, and e.g., 
${\bf s}_i = \langle r_i \rangle_0$ and 
${\bf r}_i^{{\rm N}}$ is the average and 
native position vector of the $i^{{\rm th}}$ monomer, respectively.
Let 
$\chi(t-t')^{-1}$
be the symmetric $2 \times 2$ matrix of monomer
correlations 
$\chi(t-t')^{-1} =
\langle \delta {\bf R}(t) \cdot \delta {\bf R}(t')^T \rangle_0/a^2$, 
with $\delta{\bf R}(t) = {\bf R}(t) - {\bf S}$:
\begin{equation}\label{eq:chi}
\chi(|t-t'|)^{-1} = 
\left[
\begin{array}{cc}
G_{ii} & G_{ij}(|t - t'|) \\
G_{ij}(|t-t'|) & G_{jj}
\end{array}
\right]
\end{equation}
To construct $\chi(|t-t'|)$, we have denoted the equal time correlation 
function by the static correlations $G_{ii}$ and $G_{jj}$, and used
symmetry of the correlations $G_{ij}(t) = G_{ji}(t)$.

Since this is a Gaussian process, the correlation function 
$\langle \rho_i(t) \rho_j(t) \rangle_0$ 
can be written as 
\begin{equation}\label{eq:c1}
\langle \rho_i(t) \rho_j(t') \rangle_0 = 
\int_{{\bf r}_i,{\bf r}_j} 
\! \! \! \! \!
\rho_i({\bf r}_i) \rho_j({\bf r}_j)
\tilde{P}(\delta {\bf r}_i, \delta {\bf r}_j;|t-t'|),
\end{equation}
where
\begin{eqnarray}\label{eq:pgaussian}
\lefteqn{\tilde{P}(\delta {\bf r}_i, \delta {\bf r}_j;|t-t'|)  =  
\frac{3}{2 \pi a^2} (\det \chi(|t-t'|))^{3/2}}  \hspace{1cm} \nonumber \\
& & \times  
\exp \left[ 
- \frac{3}{2 a^2} \delta {\bf R}^T \cdot \chi(|t-t'|) \cdot \delta {\bf R}
\right].
\end{eqnarray}

In the rest of the derivation, we suppress the time dependence to 
simplify notation.
Substituting Eq.(\ref{eq:loc_ndens}) for $\rho({\bf r}_i)\rho({\bf r}_j)$ and 
shifting the integration variables to $\delta {\bf R}$ 
(${\bf R} = \delta {\bf R} + {\bf S}$) leads to
\begin{eqnarray}\label{eq:c2}
\lefteqn{\langle \rho_i(t) \rho_j(t) \rangle_0 =
\frac{3}{2 \pi a^2} (\det \chi)^{3/2} 
\exp \left[ -\frac{3 \alpha^{{\rm N}}}{2 a^2} ({\bf R}^{{\rm N}} - {\bf S})^2 \right]}  
\hspace{.5cm}\nonumber \\
& & 
\times
\int_{\delta {\bf r}_i,\delta {\bf r}_j}
\! \! \! \! \!
\exp \left[
-\frac{3}{2 a^2} \delta {\bf R} \cdot \chi' \cdot \delta {\bf R} + 
\frac{3}{a^2} {\bf J} \cdot \delta {\bf R}
\right],
\end{eqnarray}
where
\begin{equation}\label{eq:chiprime}
\chi' = \chi + \alpha {\bf I}
\hspace{1cm}
I = 
\left[
\begin{array}{cc}
1 & 0 \\
0 & 1
\end{array}
\right]
\end{equation}
\begin{equation}
{\bf J} = 
{\bf R}^{{\rm N}} - {\bf S}.
\end{equation}
Performing this integral by completing the square in the exponent gives
\begin{eqnarray}\label{eq:c3}
\langle \rho_i(t) \rho_j(t) \rangle_0 & = & 
\left[ \frac{\det \chi}{\det \chi'}\right]^{3/2} 
\exp \left[ -\frac{3 \alpha^{{\rm N}}}{2 a^2} {\bf J}^2 \right] \nonumber \\
& \times & 
\exp \left[ \frac{3 (\alpha^{{\rm N}})^2}{2 a^2} 
{\bf J}^T \cdot \chi'^{-1} \cdot {\bf J}
\right].
\end{eqnarray}

To simplify this equation, it is convenient to re-write $\chi'$ as
\begin{equation}
\chi' =  \chi M
\end{equation}
where 
\begin{eqnarray}
M & = & ( I + \alpha^{{\rm N}} \chi^{-1}) \nonumber \\
  & = & 
\left[
\begin{array}{cc}
1+\alpha^{{\rm N}} G_{ii} & \alpha^{{\rm N}} G_{ij}(|t-t'|) \\
 \alpha^{{\rm N}} G_{ij}(|t-t'|) & 1+\alpha^{{\rm N}} G_{ij}
\end{array}
\right].
\end{eqnarray}
Then we can express $\chi'^{-1}$ as
\begin{equation}
\chi'^{-1} = M^{-1} \chi^{-1}.
\end{equation}
Now since
\begin{eqnarray}
I & = & M^{-1}M \\
  & = & M^{-1}(I + \alpha^{{\rm N}} \chi^{-1})\\
  & = & M^{-1} + \alpha^{{\rm N}} M^{-1} \chi^{-1}\\
  & = & M^{-1} + \alpha^{{\rm N}} \chi'^{-1},
\end{eqnarray}
we can express $\chi'^{-1}$ as
\begin{equation}
\chi'^{-1} = \frac{1}{\alpha^{{\rm N}}}( I - M^{-1}).
\end{equation}

Consequently, the combined argument the exponential factors in 
Eq.(\ref{eq:c3}) is
\begin{equation}
\frac{3(\alpha^{{\rm N}})^2}{2 a^2} 
( {\bf J}^T \cdot \chi'^{-1} \cdot {\bf J} - 
\frac{1}{\alpha^{{\rm N}}} {\bf J}^2 )  = 
- \frac{3 \alpha^{{\rm N}}}{2 a^2} {\bf J}^T \cdot M^{-1} \cdot {\bf J}.
\end{equation}
The determinate factor can be simplified using
\begin{equation}
\det \chi' = \det \chi \det M.
\end{equation}
Thus, Eq.(\ref{eq:c3}) becomes
\begin{equation}
\langle \rho_i(t) \rho_j(t) \rangle_0 = 
(\det M)^{-3/2}
\exp \left[-
\frac{3\alpha^{{\rm N}}}{2 a^2} {\bf J}^T \cdot M^{-1} \cdot {\bf J}
\right],
\end{equation}
which is  Eq.(\ref{eq:rhocorr}).


\newpage

\begin{table}
\begin{centering}
\begin{tabular}{|c|cccc|}
    &    $TS_1$ & $TS_2$ & $TS_3$ & $TS_4$ \\
\hline
$k_0 \times 10^{3} \: [\sigma]$ & 1.1  & 1.2 & 2.4 & 0.24\\
$k_0 \: [\mu \mathrm{s}^{-1}]$ & 6.8 & 7.6 & 15 & 1.5 \\
\end{tabular}
\caption
[Prefactors for $\lambda$-repressor at the folding transition temperature.] 
{
Prefactor $k_0 = |\kappa^{\star}|/2\pi$ in units of $\sigma = 3 D_0/a^2$ and
in $\mu \mathrm{s}$
for $\lambda$-repressor at the folding transition temperature, $T_{\mathrm{f}}$.  
The parameters are: $l = 5 \,a$ ($g = 0.8$), $\alpha = 0.5$, 
$D_0 = 3 \times 10^{-6} \mathrm{cm}^2/\mathrm{s}$, $\gamma_I/\gamma_0 = 9.1$.
}
\label{tab:k0_tf}
\end{centering}
\end{table}

\begin{table}
\begin{centering}
\begin{tabular}{|ccccc|}
$\ln K$ & $k_B T/\epsilon_0$ & $\Delta F^{\dagger}/k_B T$ 
   & $k_0 \: [\mu \mathrm{s}^{-1}]$ &
   $\tau \: [\mu \mathrm{s}]$  \\
\hline
0.0 &1.44& 5.1 & 7.6 & 23 \\
7.8 &1.30 & 3.4 & 2.3 & 13\\
\end{tabular}
\caption
[Kinetic parameters at two stabilities.]
{
Calculated kinetic parameters at two stabilities
$\ln K = -\beta( F_N - F_G)$ where $F_N$ and $F_G$ are the free energy of the
native and globule states:  contact energy $\epsilon_0/k_B T$, 
barrier height $\Delta F^{\dagger}/k_B T$,
corresponding folding prefactor $k_0 = |\kappa^{\star}|/2\pi$, and folding time $\tau = k^{-1}$.
Other parameters are fixed: $l = 5 \,a$ ($g = 0.8$), $\alpha = 0.5$, 
$D_0 = 3 \times 10^{-6} \mathrm{cm}^2/\mathrm{s}$, $\gamma_I/\gamma_0 = 9.1$.
}
\label{tab:k0_stab}
\end{centering}
\end{table}

\begin{table}
\begin{centering}
\begin{tabular}{|cccc|}
$D_0^{\mathrm{H_2O}} \times 10^{6} \: [\mathrm{cm}^2/\mathrm{s}]$ 
   & $\tau_{\mathrm{flip}}^{-1} \: [\mathrm{ns}^{-1}]$ 
   & $\gamma_I/\gamma_0$ & $k_0 \: [\mu \mathrm{s}^{-1}]$ \\
\hline
3 & 3 & 9.1 & 7.6\\
3 & 10 & 2.5 & 8.7 \\
9.1 & 3 & 28 & 18 \\
9.1 & 10 & 8.2 & 23\\
\end{tabular}
\caption
[Internal friction and prefactors for different 
values of the diffusion coefficient and isomerization rate.]
{
Internal friction and prefactor (from $TS_2$) for $\lambda$-repressor 
for different
values of the  monomer diffusion coefficient in water $D_0^{\mathrm{H_2O}}$
and isomerization rate $\tau_{\mathrm{flip}}^{-1}$. 
$D_0^{\mathrm{H_2O}} = 9.1 \times 10^{-6} \, \mathrm{cm}^2/\mathrm{s}$ 
is the diffusion coefficient of free alanine 
\protect{\cite{creighton:93}} 
and 
$\tau_{\mathrm{flip}}^{-1} = 10 \, \mathrm{ns}^{-1}$ is the calculated
isomerization rate for di-alanine 
\protect{\cite{pgb:cd:dc:00}}
.
The parameters in the first line are considered the best estimates.  
The persistence length is $l = 5 \,a$ ($g = 0.8$), the width of the Gaussian measure
of the native density is $\alpha = 0.5$, and the temperature is the folding 
transition temperature.
}
\label{tab:k0_vary}
\end{centering}
\end{table}

 
\end{document}